\documentclass[sigplan,screen]{acmart}
\usepackage{amsmath,bbm}
\usepackage{pgfplots}
\usepackage{latexsym,epic,eepic}
\usepackage{multirow}
\usepackage{url}
\usepackage{microtype}
\usepackage{hyperref}
\usepackage{btran}
\usepackage{booktabs}
\usepackage[all]{nowidow}
\usepackage{listings}
\usepackage{tikz}
\usepackage{stfloats}
\usepackage{algorithm}
\usepackage{algpseudocode}
\usepackage{wrapfig}
\algrenewcommand\textproc{}
\usepackage{stmaryrd}
\usetikzlibrary{shapes}
\usetikzlibrary{automata,positioning}
\newcommand{\setalglineno}[1]{%
	\setcounter{ALG@line}{\numexpr#1-1}}

\makeatother

\newenvironment{to-do}
{ \rule{1ex}{1ex}\hspace{\stretch{1}} \bfseries}
{ \hspace{\stretch{1}}\rule{1ex}{1ex} \vspace{1ex}}

\include{comm-anthony}
\include{comm}

\definecolor{darkgreen}{rgb}{0.0, 0.5, 0.0}

\newcommand\shortlong[2]{#1}

\newcommand{\tool}{CertiStr}

\usepackage{color}
\shortlong{}{\pagestyle{plain}}

\newcommand{\fixes}{\textbf{\textcolor{blue}{fixes}}}
\newcommand{\assumes}{\textbf{\textcolor{blue}{assumes}}}
\newcommand{\shows}{\textbf{\textcolor{blue}{shows}}}

\AtBeginDocument{%
	\providecommand\BibTeX{{%
			\normalfont B\kern-0.5em{\scshape i\kern-0.25em b}\kern-0.8em\TeX}}}

\setcopyright{rightsretained}
\acmPrice{}
\acmDOI{10.1145/3497775.3503691}
\acmYear{2022}
\copyrightyear{2022}
\acmSubmissionID{poplws22cppmain-p42-p}
\acmISBN{978-1-4503-9182-5/22/01}
\acmConference[CPP '22]{Proceedings of the 11th ACM SIGPLAN International Conference on Certified Programs and Proofs}{January 17--18, 2022}{Philadelphia, PA, USA}
\acmBooktitle{Proceedings of the 11th ACM SIGPLAN International Conference on Certified Programs and Proofs (CPP '22), January 17--18, 2022, Philadelphia, PA, USA}

\begin{document}
\title{\tool: A Certified String Solver}

\author{Shuanglong Kan}
\affiliation{%
	\department{Department of Computer Science}
	\institution{Technische Universit{\"a}t Kaiserslautern}
	\city{Kaiserslautern}
	\country{Germany}
}
\email{shuanglong@cs.uni-kl.de}

\author{Anthony Widjaja Lin}
\affiliation{%
	\department{Department of Computer Science}
	\institution{Technische Universit{\"a}t Kaiserslautern \& MPI-SWS}
	\city{Kaiserslautern}
	\country{Germany}
}
\email{lin@cs.uni-kl.de}

\author{Philipp R\"ummer}
\affiliation{
	\department{Department of Information Technology}              
	\institution{Uppsala University}            
	\city{Uppsala}
	\country{Sweden}
}
\email{philipp.ruemmer@it.uu.se} 
\author{Micha Schrader}
\affiliation{%
	\department{Department of Computer Science}
	\institution{Technische Universit{\"a}t Kaiserslautern}
	\city{Kaiserslautern}
	\country{Germany}
}
\email{schrader@rhrk.uni-kl.de}

\begin{abstract}
    Theories over strings are among the most heavily researched logical theories in
the SMT community in the past decade, owing to the error-prone nature of string
manipulations, which often leads to security vulnerabilities (e.g. cross-site
scripting and code injection). The majority of the existing
decision procedures and solvers for these theories are themselves intricate;
they are complicated algorithmically, and also have to deal with a very
rich vocabulary of operations. This has led to a
plethora of bugs in implementation, which have for instance been discovered 
through fuzzing. 

In this paper, we present {\tool}, a
certified implementation of a string constraint solver for the theory of strings 
with concatenation and regular constraints. 
\tool~aims to solve string constraints using a forward-propagation algorithm based 
on symbolic representations of regular constraints as symbolic automata, which
returns three results: \emph{sat}, \emph{unsat}, and \emph{unknown}, and  
is guaranteed to terminate for the string constraints
whose concatenation dependencies are acyclic.
The implementation has been developed and 
proven correct in Isabelle/HOL, through which an effective solver in OCaml was generated.  
We demonstrate the effectiveness and efficiency of \tool~against the standard Kaluza 
benchmark,
in which 
80.4\% tests are in the string constraint fragment of \tool. 
Of these 80.4\% tests,
\tool~can solve 83.5\% (i.e. \tool~returns \emph{sat} or \emph{unsat}) within 60s. 

\end{abstract}




\maketitle

\section{Introduction}
\label{sec-introduction}

Strings are among the most fundamental and commonly used data types in virtually all modern programming languages, especially with the rapidly growing popularity of dynamic languages, including JavaScript and Python. Programs written in such languages often implement security-critical infrastructure, for instance web applications, and they tend to process data and code in string representation by applying built-in string-manipulating functions; for instance, to split, concatenate, encode/decode, match, or replace parts of a string. Functions of this kind are complex to reason about and can easily lead to programming mistakes. In some cases, such mistakes can have serious consequences, e.g., in the case of web applications, cross-site scripting (XSS) attacks can be used by a malicious user to attack both the web server or the browsers of other users.

One promising research direction, which has been intensively pursued in the SMT
community in the past ten years, is the development of SMT solvers for theories
of strings (dubbed \emph{string solvers}) including
Kaluza
\cite{DBLP:conf/sp/SaxenaAHMMS10},
CVC4 \cite{cvc4}, Z3 \cite{Z3}, Z3-str3 \cite{Z3-str3}, Z3-Trau \cite{Z3-trau},
S3P \cite{TCJ16}, OSTRICH \cite{CHL+19}, SLOTH \cite{HJLRV18}, and Norn 
\cite{Abdulla14}, to name only a few. Such solvers are highly optimized
and find application, among
others, in bounded model checkers and symbolic execution tools \cite{DBLP:conf/cav/CordeiroKKST18,DBLP:conf/tacas/NollerPFLV19},
but also in tools tailored to verification of the security properties
 \cite{DBLP:conf/fmcad/BackesBCDGLRTV18}.

It has long been observed that constraint solvers
are extremely complicated procedures, and that implementations are
prone to bugs. Defects that affect soundness or completeness are
routinely found even in well-maintained state-of-the-art tools,
both in real-world applications and through techniques like fuzzing \cite{BM20,
DBLP:conf/cav/BlotskyMBZKG18,Mansur20}.
String solvers are particularly troublesome in this context, since,
unlike other SMT theories, the theory of strings requires a multitude of operations including concatenation, regular constraints, string replacement, length constraints, and many others. This inherent intricacy is also reflected
in the recently developed SMT-LIB standard for strings
(\url{http://smtlib.cs.uiowa.edu/theories-UnicodeStrings.shtml}). Many of the string solvers that arose in
the past decade relied on inevitably intricate decision procedures, which
resulted in subtle bugs in implementations themselves even among the mainstream
string solvers (e.g. see \cite{BM20,Mansur20}). This implies the unfortunate
fact that one \emph{cannot} blindly trust the answer provided by string
solvers. 

\paragraph{Contributions.}
In this paper, we present \tool,  the \emph{first certified implementation of a 
string solver} for the standard theory of strings with concatenation and 
regular constraints (a.k.a. word equations with regular constraints
\cite{Makanin,Gut98,J16,Diekert}). 
\tool~aims to solve string constraints by means of the so-called 
\emph{forward-propagation algorithm}, which relies on a simple idea of
propagating regular constraints in a \emph{forward direction} in order to derive
contradiction, or prove the absence thereof (see Section \ref{sec:example} for 
an example). Similar ideas are already present in abstract interpretation of 
string-manipulating programs (e.g. see
\cite{DBLP:journals/fmsd/YuABI14,Min05,SLOG}), but not yet at the level of
string solvers, i.e., which operate on string constraints. We have proven in
Isabelle/HOL~\cite{nipkow2002isabelle} some crucial properties of the forward-propagation algorithm
and more specifically with respect to our implementation \tool:
(i)~[TERMINATION]: the algorithm terminates on string constraints without 
cyclic concatenation dependencies, (ii)~[SOUNDNESS]: 
\tool~
is sound for \emph{unsatisfiable} results (if forward-propagation
detects inconsistencies in a constraint, then
the constraint is indeed unsatisfiable), and
(iii)~[COMPLETENESS]: 
\tool~ is complete for constraints satisfying 
the \emph{tree property} (i.e., in which on the right-hand side of each 
equation every variable appears at most once). Here
completeness amounts to the fact that, if forward-propagation does not
detect inconsistencies for a constraint satisfying the tree
property, then the constraint is indeed  \emph{satisfiable.}
For the constraints that do not satisfy the \emph{tree property} and 
\tool~cannot decide them as \emph{unsatisfiable}, \tool~returns \emph{unknown}.

In order to facilitate the propagation of regular constraints, 
we also implement a certified library for Symbolic Non-deterministic Finite 
Automata (s-NFA), which contains various automata operations, such as the concatenation and
product of two s-NFAs, the language emptiness checking for s-NFAs, among others.
s-NFAs --- as introduced by Veanes et al. \cite{symbolic-transducer} (see
\cite{DV21} for more details) ---
are different from classic Non-deterministic Finite Automata (NFA) by
allowing transition labels to be a set of characters in the (potentially
infinite) alphabet, represented by an element of a boolean algebra (e.g. the
interval algebra or the BDD algebra), instead of a single character. s-NFAs are 
especially crucial for an efficient implementation of automata-based string
solving algorithms and many string processing algorithms 
\cite{mona-secret,symbolic-transducer,DV21,CHL+19,Abdulla14}, for which
reason our certified implementation of an s-NFA library is of independent
interests. 

Last but not least, we have automatically generated a verified implementation
\tool~in Isabelle/HOL, which we have extensively evaluated against the
standard string solving benchmark from Kaluza \cite{DBLP:conf/sp/SaxenaAHMMS10}
with around 38000 string constraints. For the first time, we demonstrate
that the simple forward-propagation algorithm in fact performs surprisingly 
well even compared to other highly optimized solvers (which are not verified
implementations). In particular, we show that the majority of these constraints
(83.5\%) are solved by \tool; for the rest, the tool either
returns \emph{unknown}, or times out. Moreover, \tool~terminates within 60 seconds on 98\% of
the constraints, witnessing its efficiency.


To summarize, our contributions are:

\begin{itemize}
	\item we developed in Isabelle/HOL the tool \tool, the first certified 
        implementation of a string solver.
	\item we implemented the first certified symbolic automata library,
        which is crucial for an efficient implementation of
        many string processing applications \cite{DV21}.
	\item \tool~was evaluated over Kaluza benchmark 
        \cite{DBLP:conf/sp/SaxenaAHMMS10}, with around 38000 tests. 
        In this benchmark, 83.5\% of the tests can be solved with the results \emph{sat} or \emph{unsat}.
        Moreover, the solver terminates within 60 seconds on 98\% of  the tests,
        witnessing the surprising competitiveness of the simple 
        forward-propagation algorithm against more complicated algorithms.
\end{itemize}


\section{Motivating Example}
\label{sec:example}

We start by illustrating the decision procedure implemented by \tool. \tool~uses the so-called
\emph{forward-propagation algorithm} for solving satisfiability of
string theory with concatenation and regular membership constraints.
To illustrate the algorithm, consider the formula:
\begin{align}
  \mathit{domain} &~\in~ \texttt{/[a-zA-Z.]+/}
                                                    \label{eq:1}
  \\
  \hspace*{-1.3em}
  \mbox{}\wedge \makebox[8ex][r]{$\mathit{dir}$, $\mathit{file}$} &~\in~ \texttt{/[a-zA-Z0-9.]+/}
                                                    \label{eq:2}
  \\
  \hspace*{-1.3em}
  \mbox{}\wedge \makebox[8ex][r]{$\mathit{path}$} & ~=~ \mathit{dir}
                                                   +
                                                   \texttt{"/"} + \mathit{file}
                                                    \label{eq:3}
  \\
  \hspace*{-1.3em}
  \mbox{}\wedge \makebox[8ex][r]{$\mathit{url}$} & ~=~ \texttt{"http://"} +
                                                   \mathit{domain} +
                                                   \texttt{"/"} + \mathit{path}
                                                    \label{eq:4}
  \\
  \hspace*{-1.3em}
  [\mbox{}\wedge \makebox[8ex][r]{$\mathit{url}$} &~\in~ \texttt{/.*<script>.*/}]
                                                    \label{eq:5}
\end{align}

The formula contains string variables (\textit{domain}, \textit{dir},
etc.), and uses both regular membership constraints and word equations
with concatenation to model the construction of a URL from individual
components. Regular expressions are written using the standard PCRE
syntax (\url{https://perldoc.perl.org/perlre}). Each equation should be read as an assignment
of the right-hand side term to the left-hand side variable, and is
processed in this direction.  We initially ignore the constraint~\eqref{eq:5}.

The forward-propagation algorithm propagates regular membership
constraints and derives new constraints for the variable on the
left-hand side of equations. In this example, the algorithm
starts with equation~\eqref{eq:3} and propagates
constraint~\eqref{eq:2}, resulting in the new constraint
\begin{equation}
  \mathit{path} ~\in~ \texttt{/[a-zA-Z0-9.]+\textbackslash/[a-zA-Z0-9.]+/}
\end{equation}
We can propagate this new constraint further, using
equation~\eqref{eq:4} and together with constraint~\eqref{eq:1}, deriving:
\begin{multline}
  \mathit{url} ~\in~ \texttt{/http:\textbackslash/\textbackslash/[a-zA-Z.]+}\\
  \texttt{\textbackslash/[a-zA-Z0-9.]+\textbackslash/[a-zA-Z0-9.]+/}
  \label{eq:7}
\end{multline}

At this point, no more propagations are possible.
Note that forward-propagation will terminate, in
general, whenever no cyclic dependencies exist between the equations,
which is the fragment of formulas we consider.

Still ignoring \eqref{eq:5}, we can observe that forward-propagation
has not discovered any inconsistent constraints. In order to conclude
the \emph{satisfiability} of the formula from this, we need a 
meta-result about forward-propagation: we show that forward
propagation is complete for acyclic formulas in which each variable
occurs at most once on the right-hand side of equations. Since this
criterion holds for the
formula~$\eqref{eq:1}\wedge\cdots\wedge\eqref{eq:4}$, we have indeed
proven that it is satisfiable.

Consider now also \eqref{eq:5}, modelling a simple form of injection
attack. We can observe that this additional constraint is inconsistent
with the derived constraint \eqref{eq:7}, since the intersection of
the asserted regular languages is empty. Forward-propagation detects
this conflict as soon as \eqref{eq:7} has been computed: each time a
new membership constraint is derived, the algorithm will check the
consistency with constraints already assumed. Since we also show that
forward-propagation is sound w.r.t. inconsistency, the detection of a conflict immediately
implies that the formula $\eqref{eq:1}\wedge\cdots\wedge\eqref{eq:5}$
is unsatisfiable.

\medskip
In summary, the forward-propagation is defined using two main
inference steps: the post-image computation for word equations, and
a consistency check for regular constraints. Although the
example uses regular expressions, both operations can be defined more
easily through a translation to finite-state automata. To obtain a
verified string solver, the implementation of both operations has to be shown correct,
and in addition meta-results need to be derived about the soundness
and completeness of the overall algorithm.

We can also observe that classical finite-state automata, over concrete
alphabets, are cumbersome even for toy examples; regular expressions
often talk about character ranges that would require a large number of
individual transitions.  In practice, string constraints are usually
formulated over the Unicode alphabet with currently $3 \times 2^{16}$
characters, which necessitates symbolic character representation. In
our verified implementation, we therefore use a simple form of
symbolic automata~\cite{DBLP:conf/cav/DAntoniV17} in which transitions
are labelled with character intervals.


\section{String Constraint Fragment}
\label{sec-str-constraint}

In this section, we present the string constraint fragment that \tool~supports and
its semantics of satisfiability.

Let $\Sigma$ be an alphabet and
$\Sigma^*$ denote the set of words over $\Sigma$.
Let $w_1, w_2 \in \Sigma^*$ be two words,
$w_1w_2$ denotes the concatenation of the two words, i.e.,
$w_2$ is appended to the end of $w_1$ to get a new string.
We consider the following string constraint language over $\Sigma$:
\[
	c ::= x \in \mathcal{A} \mid  x = x_1+ x_2 \mid  c_1\wedge c_2,
\]
where $x$ denotes a variable ranging over $\Sigma^*$ and
$\mathcal{A}$ denotes an NFA representing
a regular language over $\Sigma$. 
The language accepted by $\mathcal{A}$ is denoted by $\mathcal{L}(\mathcal{A})$.
The semantics of a constraint $c$ is defined in an obvious way by interpreting
$x \in \mathcal{A}$ as a membership of $x$ in the language 
$\mathcal{L}(\mathcal{A})$, $x = x_1 + x_2$ as a string equation, and
$c_1\wedge c_2$ as a conjunction
of $c_1$ and $c_2$.
More precisely, given a function $\mu$ mapping each variable in the constraint
$c$ to a string in $\Sigma^*$, we say that $\mu$ \emph{satisfies} $c$ if: (i)
$\mu(x) \in \mathcal{L}(\mathcal{A})$ for each constraint $x \in \mathcal{A}$ in
$c$, and (ii) $\mu(x) = \mu(x_1)\mu(x_2)$ for each constraint $x = x_1 + 
x_2$. The constraint $c$ is satisfiable if one such \emph{solution} $\mu$ for
$c$ exists.

%
%
%
%

Our string constraint language is as general as word equations with regular
constraints \cite{Makanin,Gut98,J16,Diekert}, which forms the basis of the
recently published Unicode string constraint language in SMT-LIB 2.6, and already
makes up the bulk of existing string constraint benchmarks.
Notice that our restriction to string equation of the form $x = x_1 + x_2$
is not a real restriction since general string constraints can be obtained by 
means of desugaring.
For instance, the concatenation of more than two variables, 
like $x=x_1 + x_2 + x_3$ can be desugared as two concatenation constraints:
$(1)~x' =x_1+ x_2;x=x'+ x3$.
Moreover, a subset of string constraints with length functions (called monadic length) can also be translated
into this fragment with regular  constraints.
For instance, $|x| \leq 6$, where $|x|$ denotes the length of the word of $x$, can be translated to a regular membership constraint as $x \in \mathcal{\mathcal{A}}$, where $\mathcal{\mathcal{A}}$ is an NFA that accepts any word whose length is at most $6$.
Recent experimental evidence shows that a substantial portion of length 
constraints that appear in practice are essentially monadic \cite{HLRW20}.
There are also some string constraints using disjunction ($\vee$).
For instance, $c_1\vee c_2$, where $c_1$ and $c_2$ are string constraints in our fragment.
This also can be solved with \tool~by first solving $c_1$ and $c_2$ separately,
and then checking whether one of them is satisfiable.

\begin{figure*}
	\vspace{3mm}
	\includegraphics[scale=0.5]{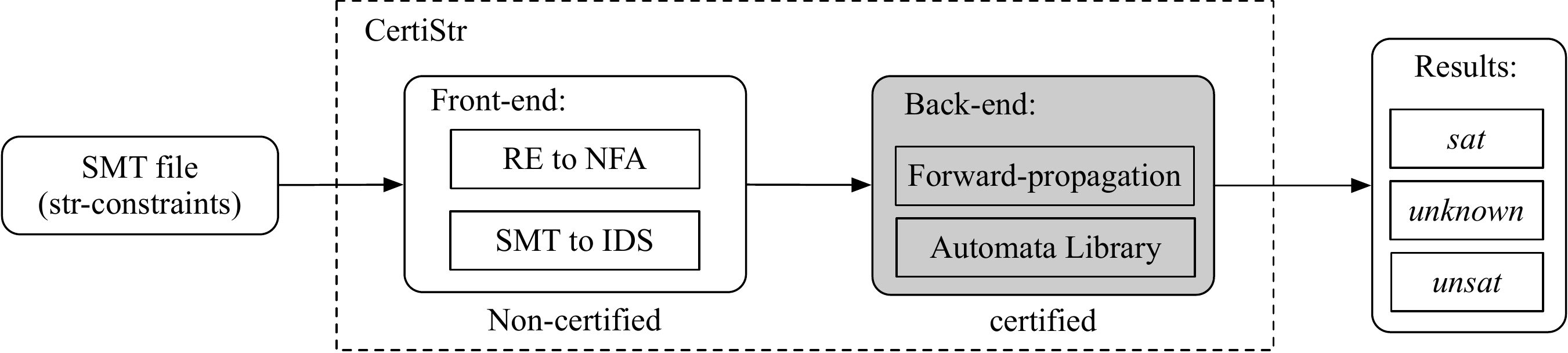}
	\caption{The framework of the certified string solver}
	\label{fig-workflow}
\end{figure*}

Figure \ref{fig-workflow} 
shows the framework of \tool.
The input of the solver is an SMT file
with string constraints.
The tool has two parts:
(1) the front-end (non-certified) and
(2) the back-end (certified).
The front-end parses the SMT file,
which will first translate all Regular Expressions (RE) to NFAs
and desugar some string constraints that are not in the fragment,
and then translate the string constraints
to the Intermediate Data Structures (IDS),
which are the inputs of the back-end.
The IDS contains three parts: 
(1) a set $S$ of variables that are used in the string constraints, 
(2) a concatenation constraint map $\mathit{Concat}$, which maps
	 variables in $S$ to the set of their concatenation constraints, and
(3) a regular constraint map $\mathit{Reg}$, which maps each variable in $S$ to its regular membership constraints.

The back-end contains two parts: (1) the Automata Library, which contains a collection of automata operations, such as the product and concatenation of two NFAs, 
and (2) 
a forward-propagation to check whether the string constraints are satisfiable.
The results of the certified string solver 
can be
(1) \emph{sat} the string constraint is satisfiable,
(2) \emph{unsat} the string constraint is unsatisfiable, and
(3) \emph{unknown} the solver cannot decide whether
the string constraint is satisfiable or not.
Note that, \emph{unknown} does not mean that \tool~is non-terminating.
It means the results after executing the forward-propagation for 
string constraints are not sufficient to decide their satisfiability.

\section{Forward-propagation of String Constraints}  
\label{sec-forward-analysis}
 In this section, we present the algorithm of the forward-propagation
 and the properties proven for it.
\subsection {The Algorithm of Forward-propagation}
We use the following example of string constraints to illustrate
our forward-propagation analysis.
\begin{example}
	\label{example-str-constraints}
	$x_1\in \mathcal{A}_1\wedge x_2\in \mathcal{A}_2\wedge x_3\in \mathcal{A}_3\wedge x_5=x_3+ x_4\wedge x_3 = x_1+ x_2$.
\end{example}

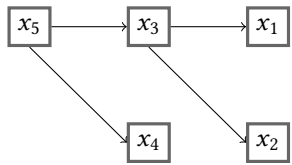
\begin{figure}[ht]
	\centering
\begin{tikzpicture}[
	squarednode/.style={rectangle, draw=black!60, very thick, minimum size=5mm},
	]
	\node[squarednode]      (x5)                              {$x_5$};
	\node[squarednode]      (x3)       [right=of x5] {$x_3$};
	\node[squarednode]      (x4)       [below=of x3] {$x_4$};
	\node[squarednode]      (x1)       [right=of x3] {$x_1$};
		\node[squarednode]      (x2)       [below=of x1] {$x_2$};
	
	\draw[->] (x5.east) -- (x3.west);
	\draw[->] (x5.south) -- (x4.west);
	\draw[->] (x3.east) -- (x1.west);
	\draw[->] (x3.south) -- (x2.west);
\end{tikzpicture}
\caption{Dependence graph of variables in Example \ref{example-str-constraints}} \label{fig-example-str-constraints}
\end{figure}

Example \ref{example-str-constraints} contains five variables ``$x_1,x_2,x_3,x_4,x_5$'' and two concatenation constraints
``$x_5=x_3+ x_4\wedge x_3 = x_1+ x_2$''.
We can view these concatenation constraints as a dependence graph shown in 
Figure \ref{fig-example-str-constraints}.
For a concatenation constraint $x_k = x_i + x_j$,
it yields two dependence relations: 
$x_k\mapsto x_i$ and $x_k\mapsto x_j$ ($a\mapsto b$ denotes $a$ depends on $b$).
This means that 
before propagating the concatenation of the regular constraints of  
$x_i$ and $x_j$ to $x_k$,
we need to first compute the propagation  to $x_i$ and $x_j$.
Let the initial regular constraint of $x_k$ be $\mathcal{A}_k$.
Let the regular constraints of $x_i$ and $x_j$ be $\mathcal{A}_i$ and $\mathcal{A}_j$, respectively.
The new regular constraint of $x_k$ after the propagation can be refined to
an automaton, whose language is $\mathcal{L}(\mathcal{A}_k)\cap(\mathcal{L}(\mathcal{A}_i)+\mathcal{L}(\mathcal{A}_j))$,
where $\mathcal{L}_1+\mathcal{L}_2\triangleq \{w_1w_2\mid  w_1\in \mathcal{L}_1 \land w_2\in \mathcal{L}_2\}$.
Here, the term ``refine" means narrowing the regular constraint of a variable
by propagating the concatenation constraints.

The forward-propagation repeats the following computation
 until all variables are refined:
 
 \begin{itemize}
 	\item detect a set of variables such that for each variable in the set, all its dependence variables have already been refined.
 	\item refine the regular constraints of the variables detected in the last step with respect to their concatenation constraints.
 \end{itemize}

We illustrate the idea of the forward-propagation with
Example \ref{example-str-constraints}. The first iteration of the forward-propagation
will detect the set of variables $\{x_1,x_2,x_4\}$ since they have no dependence variables and
we do not need to refine their regular constraints.
The second iteration of the forward-propagation will detect the set of variables: $\{x_3\}$,
as $x_3$ only depends on the variables in $\{x_1,x_2,x_4\}$.
Its regular constraint will be refined to an automaton, whose language is $\mathcal{L}(\mathcal{A}_3)\cap(\mathcal{L}(\mathcal{A}_1)+\mathcal{L}(\mathcal{A}_2))$
after the propagation for the constraint $x_3=x_1+ x_2$.
Assume this new regular constraint as $\mathcal{A}_3'$.
The third iteration detects the variable $x_5$,
whose  regular constraint will be refined to an automaton, whose language is $\Sigma^*\cap (\mathcal{L}(\mathcal{A}_3')+\Sigma^*)$.
Note that we do not specify the initial regular constraints of $x_4$ and $x_5$ in the string constraint and
therefore their initial regular constraints are $\Sigma^*$ by default.

Algorithm \ref{alg-forward-analysis} shows
the forward-propagation algorithm.
It contains three procedures: $\mathbf{Forward\_Prop}$, $\mathbf{Ready\_Set}$, and
$\mathbf{Var\_Lang}$.
The procedure $\mathbf{Forward\_Prop}$ is the entry point of the algorithm and has 
three parameters:
(1) $S$ is the set of variables used in the string constraints,
for instance, in Example \ref{example-str-constraints}, there are five variables: $S = \{x_1,x_2,x_3,x_4 , x_5\}$;
(2) $\mathit{Concat}: S\rightharpoonup 2^{(S\times S)}$ is a partial map 
which maps a variable $x$ to a set of pairs of variables.
For each pair $(x_1,x_2)$ in the set, there is a concatenation 
constraint: $x = x_1+ x_2$;
(3) $\mathit{Reg}:S\rightarrow \mathcal{N}$ is a map from variables to automata. These automata denote
the variables' initial regular constraints. Here $\mathcal{N}$ denotes the set of NFAs.

\begin{algorithm}
	\caption{The algorithm of forward-propagation}
	\label{alg-forward-analysis}
		\begin{algorithmic}[1]
			\Procedure{$\mathbf{Forward\_Prop}$}{$S, \mathit{Concat}, \mathit{Reg}$}~~
			\State $R\gets \emptyset$
			\While{$S\neq\emptyset$}
			\State $C$ $\gets$ \Call{$\mathbf{Ready\_Set}$}{$S, \mathit{Concat}, R$}
			\State $\mathit{Reg}$ $\gets$ \Call{$\mathbf{Var\_Lang}$}{$C, \mathit{Concat}, \mathit{Reg}$}
			\State $S$ $\gets$ $S~-~C$; $R$ $\gets$ $R~\cup~C$
			\EndWhile
			\State \textbf{return} $\mathit{Reg}$
			\EndProcedure
			\Procedure{$\mathbf{Var\_Lang}$}{$C, \mathit{Concat}, \mathit{Reg}$}
			\For{each $v\in C$}
			\State $\mathcal{A}$ $\gets$ $(\mathit{Reg}~v)$
			\For{each $(v_1,v_2)\in (\mathit{Concat}~v)$}
			\State $\mathcal{A'}\gets \mathbf{NFA\_concat}~(\mathit{Reg}~v_1)~(\mathit{Reg}~v_2)$
			\State $\mathcal{A}$ $\gets$ $\mathbf{NFA\_product}~\mathcal{A}~~\mathcal{A}'$
			\EndFor
			\State $\mathit{Reg}~\gets~\mathit{Reg}~[v\mapsto \mathcal{A}]$ 
			\EndFor
			\State \textbf{return} $\mathit{Reg}$ 
			\EndProcedure
		\end{algorithmic}
		\begin{algorithmic}[1]
			\setalglineno{22}
			\Procedure{$\mathbf{Ready\_Set}$}{$S, \mathit{Concat}, R$}
			\State  $C$ $\gets$ $\emptyset$
			\For{each $v\in S$}
			\State $D$ $\gets$ $\emptyset$
			\For{each $(v_1,v_2)\in (\mathit{Concat}~v)$}
			\State $D$ $\gets$ $D~\cup~\{v_1,v_2\}$
			\EndFor
			\If{$D~\subseteq R$}
			\State $C~\gets~C~\cup~\{v\}$
			\EndIf
			\EndFor
			\State \textbf{return} $C$
			\EndProcedure
		\end{algorithmic}
\end{algorithm}

In the while loop of the procedure $\mathbf{Forward\_Prop}$, the computation
iteratively refines the regular constraints of variables by propagating
the regular constraints of variables on the right hand side of the concatenation constraints.
The first step in the loop computes the set $C$ of variables whose dependence variables
are all in the set $R$ (computed by the procedure $\mathbf{Ready\_Set}$). $R$ is initialized empty and it denotes the set of variables that have already been refined by the previous loops.
The second step in the loop
updates the regular constraints of the variables in $C$
stored in $\mathit{Reg}$ by propagating the concatenation constraints in $\mathit{Concat}$ (computed by $\mathbf{Var\_Lang}$).

At the end of the loop body, the set $S$ is updated by removing the variables in $C$ as they have already
been refined and these removed variables are added to the set $R$, i.e., they have already been refined.

The procedure $\mathbf{Ready\_Set}$ computes the set of variables that are only
dependent on the variables in $R$.
For each variable $v$, it stores all its dependence variables in $D$ and 
then checks whether $D$ is a subset of $R$.
If $D$ is a subset of $R$ then $v$ is moved  to $C$ as all its dependence variables
have been refined.

The procedure $\mathbf{Var\_Lang}$ refines the regular constraints of variables in $C$.
In the inner loop of $\mathbf{Var\_Lang}$, 
it traverses all concatenation constraints $v=v_1+ v_2$ and
refines the regular constraint of $v$ by propagating the concatenation of the regular constraints of $v_1$ and $v_2$.
$\mathit{Reg}~[v\mapsto \mathcal{A}]$ means updating the regular constraint of $v$ in $\mathit{Reg}$ to $\mathcal{A}$.
The functions $\mathbf{NFA\_concat}$ and $\mathbf{NFA\_product}$
denote constructing the concatenation and product of two NFAs, respectively.
These two functions will  be introduced in Section \ref{sec-SAF}.

Assume a string constraint is stored in $S$, $\mathit{Concat}$ and $\mathit{Reg}$,
then after calling ``$\mathbf{Forward\_Prop}(S,~\mathit{Concat},~\mathit{Reg})"$,
we will get a new $\mathit{Reg}'$ such that for each variable $v$, 
$(\mathit{Reg}'~v)$ stores the refined regular constraint of $v$, which is an NFA.
We have two cases to decide whether the string constraint
is satisfiable or not:
\begin{itemize}
	\item If there exists a variable $v$ such that the language of $(\mathit{Reg}'~v)$ is empty then
	the string constraint is unsatisfiable, because the refined regular constraint of $v$ is the empty language.
	\item If for all variables $v$ such that  the language of $(\mathit{Reg}'~v)$ is not empty then 
	unfortunately we cannot decide whether the string constraint is satisfiable or not. 
	Consider the string constraint: ``(1) $y = x+ x\wedge (2)~y \in \{ab\}\wedge (3)~x \in \{a,b\}$'', 
	where $ab$, $a$, and $b$ are words, the refined regular constraints of $x$ and $y$ are both not empty. But this string constraint is not satisfiable as
	constraints (1) and (2)  require $x$ to have two different values.
	We will introduce a property, called \emph{tree property} later, such that
	if the string constraint satisfies the \emph{tree property} then it is satisfiable, otherwise
	\tool~returns \emph{unknown}.
\end{itemize}

\subsection{Theorems for  Forward-propagation}

Now we introduce the key theorem proven for the algorithm.
Firstly we present the definitions of well-formed inputs ($\mathbf{wf}$ in Figure \ref{fig-def-wf-acyclic}) and the acyclic property ($\mathbf{acyclic}$ in Figure \ref{fig-def-wf-acyclic}) for concatenation constraints,
which are the premises of the correctness theorem.

The $\mathbf{wf}$ predicate requires that all variables appearing in the domain
and codomain of $\mathit{Concat}$ should be in the set $S$, and
the domain of $\mathit{Reg}$ should be the set $S$, i.e., all
variables should have their initial regular constraints.
$S$ should be finite.
Now we consider the  definition of $\mathbf{acyclic}$, which checks whether the dependence graphs
of concatenation constraints are acyclic.
$\mathbf{Forward\_Prop}$ cannot solve string constraints that are not acyclic.
For instance, the string constraint ``$x=y_1+ y_2\wedge y_1=z_1+ x$''
does not satisfy the acyclic property,
as  $x\mapsto y_1 \mapsto x$.

\begin{figure}[t]
\begin{lstlisting}
definition |$\mathbf{wf}$| where
|$\mathbf{wf}$ $S$ $\mathit{Concat}$ $\mathit{Reg}$| = 
  |$(dom$ $\mathit{Concat}$ $\subseteq$ $S$)$\land$($dom$ $\mathit{Reg}$ = $S$)$\land$|
  |$(\forall$ $v$ $v_1$ $v_2.$ $v\in$$(dom$ $\mathit{Concat}$$)\land (v_1,v_2)\in (\mathit{Concat}$ $v)$|
      |$\longrightarrow v_1\in S$ $\land$ $ v_2\in S)$| |$\land$| finite |$S$|
    
definition |$\mathbf{acyclic}$| where
|$\mathbf{acyclic}$ $S$ $\mathit{Concat}$ $l$| = |$S = \bigcup (\mathbf{set}$ $l) \land ((l=[])$|
  |$\lor(l=s\#l'\longrightarrow (s\cap (\bigcup (\mathbf{set}$ $l')) = \emptyset)$ $\land$|
  |$ (\forall v_1$ $v_2$ $v.$ $v\in s\land (v_1,v_2)\in (\mathit{Concat}$ $v) \longrightarrow$|
  |$\{v_1,v_2\}\subseteq \bigcup(\mathbf{set}$ $l')) \land $||$(\mathbf{acyclic}$ $(S-s)$ $\mathit{Concat}$ $l')))$| 
\end{lstlisting}
\caption{The definitions of $\mathbf{wf}$ and $\mathbf{acyclic}$}
\label{fig-def-wf-acyclic}
\end{figure}
In order to specify the acyclic property, 
we arrange 
the set $S$ of variables to a list in which elements are disjoint sets of variables in $S$
and the list characterizes the dependence levels among variables.
For instance, in Example \ref{example-str-constraints},
there are $5$ variables and they can
be organized in a list: $[\{v_5\}, \{v_3\}, \{v_1,v_2,$ $v_4\}]$.
Indeed, the list characterizes the dependence levels among variables.
For instance, the variable $v_5$ only depends on 
the variables in  $\{v_3\}$ and $\{v_1,v_2, v_4\}$
and $v_3$ only depends on the variables in $\{v_1,v_2,v_4\}$ but not $\{v_5\}$.
The variables in $\{v_1,v_2,v_4\}$ do not depend on any variables.
If $\mathit{Concat}$ does not satisfies the acyclic property
then we cannot arrange all variables in such a list, as all variables in a cycle
always depend on another variable in the cycle. 
%
%

The predicate $\mathbf{acyclic}$
contains two parts.
(1) $S = \bigcup (\mathbf{set}~l)$, which means the elements in $S$ and $l$ should be exactly the same.
``$\mathbf{set}~l$'' translates the list $l$ to a set of elements in $l$ and
``$\bigcup (\mathbf{set}~l)=\{x\mid \exists y. y\in \mathbf{set}~l\land x\in y\}$''.
(2) Part 2 further has two cases.
The first case is ``$l=[]$'', i.e., empty list. In this case, the predicate is obviously true.
The second case is ``$l=s\#l'$'', i.e., it has at least one element. $s$ is the first element
and $l'$ is the tail.
In this case, firstly, it requires the elements in $s$ to be 
disjoint with the elements
in the sets of the tail $l'$ (checked by $s\cap (\bigcup (\mathbf{set}~l'))= \emptyset$).
Secondly, let  $v$ be a variable in $s$,
for any two variables $v_1$ and $v_2$ that $v$ depends on,
$v_1$ and $v_2$ can only be in the sets of the tail $l'$.
Thirdly, $\mathbf{acyclic}~(S-s)~\mathit{Concat}~l'$ checks
whether the remaining elements in $S-s$
are still acyclic w.r.t. $\mathit{Concat}$ and $l'$.
In order to check whether the variables in $S$ are acyclic w.r.t. $\mathit{Concat}$, we can specify it as ``$\exists l.~ \mathbf{acyclic}~S~\mathit{Concat}~l$''.

Now we are ready to introduce our correctness theorem (Theorem \ref{thm-forward_analysis_correct}).
The keyword \fixes~in Isabelle is used to define universally quantified variables.
The keyword \assumes~is used to specify assumptions of the theorem and
the keyword \shows~is used to specify the conclusion of the theorem.
The \shows~part is of the form
''$M \leq \mathbf{SPEC}~(\lambda~Reg'.~\Phi)$'',
which means the algorithm $M$ 
is correct with respect to the specification $\Phi$.
More precisely, the result of applying the predicate $(\lambda~Reg'.~\Phi)$
to the output of $M$ should be true.
``$\mathbf{NFA\_accept}~w~\mathcal{A}$'' checks whether $w$ is accepted by $\mathcal{A}$, which will be introduced in Section \ref{sec-SAF}.
The symbol ``$@$'' is the list concatenation operation in Isabelle. We use 
a list of elements in $\Sigma$ to represent a word.

\begin{theorem}[Correctness of $\mathbf{Forward\_Prop}$]
	\label{thm-forward_analysis_correct}
	~\\
	\begin{tabular}{rp{7.2cm}}
		\fixes& $S$ $\mathit{Concat}$ $\mathit{Reg}$\\
		\assumes & 
		1. $\mathbf{wf}~S~\mathit{Concat}~\mathit{Reg}$ \\
		& 2. $(\exists l.~$ $\mathbf{acyclic}~S~\mathit{Concat}~l)$\\
		\shows &  $\text{$\mathbf{Forward\_Prop}$}(S, \mathit{Concat}, \mathit{Reg})\leq$ $\mathbf{SPEC}(\lambda\mathit{Reg'}.$\\
		&
		  $~(\forall v~w.~\mathbf{NFA\_accept}~w~(\mathit{Reg}'~v)\longleftrightarrow$\\
		& $\quad(\mathbf{NFA\_accept}~w~(\mathit{Reg} ~v)~\land$\\
		& $\quad\quad(\forall v_1~v_2.~(v_1,v_2)\in (\mathit{Concat}~v)\longrightarrow$\\
		& $\quad\quad(\exists w_1~w_2.~w=w_1@ w_2~\land~$\\
		& $\quad\quad\mathbf{NFA\_accept}~w_1~(\mathit{Reg}' ~v_1)\land$\\
		& $\quad\quad\mathbf{NFA\_accept}~w_2~(\mathit{Reg}' ~v_2))))))$\\
	\end{tabular}
\end{theorem}

Theorem \ref{thm-forward_analysis_correct}
shows that if $S,\mathit{Concat},\mathit{Reg}$ are well-formed and acyclic then
after calling $\mathbf{Forward\_Prop}(S,\mathit{Concat},\mathit{Reg})$ we can get $\mathit{Reg}'$, which satisfies:
any
$w$ is accepted by the automaton $(\mathit{Reg}'~v)$
if and only if
\begin{enumerate}
\item $w$ is accepted by the original regular constraint of $v$, i.e., $(\mathit{Reg}~v)$, and
\item for all variables $v_1$ and $v_2$, $(v_1,v_2)\in \mathit{Concat}~v$ implies
there exist $w_1,w_2$ such that
$w=w_1@ w_2$ and
$w_i,$ for $i=1,2$, is accepted by the regular constraint $(\mathit{Reg}'~v_i)$.
\end{enumerate}

That is, in the new $\mathit{Reg}'$, the concatenation constraints are used to refine
the corresponding variables.

Moreover, in Isabelle, Theorem \ref{thm-forward_analysis_correct}
also requires us to prove the termination
of  Algorithm \ref{alg-forward-analysis}.
The termination
is proven by 
showing 
that in the procedure $\mathbf{Forward\_Prop}$, 
the number of variables in $S$ 
decreases progressively for each iteration.
Since $S$ is finite, the algorithm must terminate.

Even though Theorem  \ref{thm-forward_analysis_correct} is the key of the correctness of the forward-propagation,
it does not give us an intuition with respect to the satisfiability semantics of string constraints introduced in Section \ref{sec-str-constraint}.
In the following, we will show that our forward-propagation is sound w.r.t.
\emph{unsat} results.

\begin{figure}[t]
\begin{lstlisting}
definition |$\mathbf{sat\_str}$| where
|$\mathbf{sat\_str}$ $S$ $\mathit{Concat}$ $\mathit{Reg}$ $\mu = $|
  |$(\forall v\in S.$ $(\mu$ $v)\in \mathcal{L}(\mathit{Reg}$ $v))\land$|
   |$(\forall v~v_1~v_2.~v \in S \land (v_1,v_2)\in (\mathit{Concat}$ $v) \longrightarrow$|
	|$(\mu$ $v)=(\mu$ $v_1)@(\mu$ $v_2))$|
\end{lstlisting}
	
\begin{lstlisting}
definition |$\mathbf{tree}$| where
|$\mathbf{tree}~\mathit{Concat}$| = 
  |$(\forall v~v_1~v_2.~(v_1,v_2)\in \mathit{Concat}~v~\longrightarrow v_1\neq v_2)~\land$|
  |$(\forall v~v_1~v_2~v_3~v_4.~(v_1,v_2)\neq (v_3,v_4)~\land$| 
   |$(v_1,v_2)\in \mathit{Concat}~v\land(v_3,v_4)\in \mathit{Concat}~v$|  
	|$\longrightarrow \{v_1,v_2\}\cap\{v_3,v_4\}=\emptyset)~\land$|	
  |$(\forall v$ $v'$ $v_1$ $v_2$ $v_3$ $v_4.$ $v\neq v' \land (v_1,v_2)\in \mathit{Concat}$ $v$|
   |$\land (v_3,v_4)\in \mathit{Concat}$ $v'\longrightarrow \{v_1,v_2\}\cap\{v_3,v_4\}=\emptyset)$|
\end{lstlisting}	
\caption{The definitions of $\mathbf{sat\_str}$ and $\mathbf{tree}$}
\label{figure-def-satstr-tree}
\end{figure}

Firstly we need to formalize the semantics of \emph{sat} and \emph{unsat} for string constraints.
Let 
$\mu$ be an assignment, which is a map $S\rightarrow \Sigma^*$
from variables in $S$ to words.
We define the  predicate $\mathbf{sat\_str}$ in Figure \ref{figure-def-satstr-tree} for the semantics of the satisfiability of string constraints.

The predicate $\mathbf{sat\_str}$ contains two parts:
(1) the assignment $\mu$ should respect all regular constraints and
(2) the assignment $\mu$ should respect
all concatenation constraints.

Theorem \ref{thm-forward_sound} shows that after executing $\mathbf{Forward\_Prop}$,
we get a new map $\mathit{Reg}'$.
If there exists a variable $v$, such that $\mathcal{L}(\mathit{Reg}'~v) = \emptyset$, i.e., 
the language of the refined regular constraint of $v$ is the empty set,
then there is no assignment $\mu$ that makes the predicate ``$\mathbf{sat\_str}~S~\mathit{Concat}~\mathit{Reg}~\mu$''
true.

\begin{theorem}[Soundness for \emph{unsat} results]
	\label{thm-forward_sound} 
	~\\
	\begin{tabular}{rl}
		\fixes & $S$ $\mathit{Concat}$ $\mathit{Reg}$ \\
		\assumes & 
		1. $\mathbf{wf}~S~\mathit{Concat}~\mathit{Reg}$\\
	&	2. $(\exists l.~$ $\mathbf{acyclic}~S~\mathit{Concat}~l)$\\
		\shows 
	    & $\text{$\mathbf{Forward\_Prop}$}(S, \mathit{Concat}, \mathit{Reg})\leq \mathbf{SPEC}(\lambda\mathit{Reg}'.$\\
		& $((\exists v.~\mathcal{L}(\mathit{Reg}'~v)=\emptyset) \longrightarrow$\\
		&$ (\forall \mu.~ \neg~\mathbf{sat\_str}~S~\mathit{Concat}~\mathit{Reg}~\mu)))
		$\\
	\end{tabular}
\end{theorem}

Unfortunately, the theorem holds only for \emph{unsat} results.
As we already analyzed before,
the case $\forall v.~\mathcal{L}(\mathit{Reg}'~v)\neq \emptyset$ 
cannot make us  conclude that the string constraint is satisfiable.

In this paper, we characterize a subset of the string constraints, using a property such that
the forward-propagation's \emph{satisfiable} results are also sound.
We call it the \emph{tree property}.
The intuition behind the \emph{tree property} 
is that the dependence graph of a string constraint should
constitute a tree or forest.
The example $y=x+ x$ does not satisfy the  \emph{tree property},
because $y$ has two edges pointing to a same node $x$ in the dependence graph.
In other words, the node $x$ has indegree 2.
But a tree requires that each node has at most indegree 1.
As the map $\mathit{Concat}$ stores the dependence graphs of string constraints,
we only need to check $\mathit{Concat}$ for the tree property.
The predicate $\mathbf{tree}$ is defined in Figure \ref{figure-def-satstr-tree}.


This definition has three cases separated by the conjunction operator: 
(1) the first case 
requires that for any concatenation constraint $v=v_1+ v_2$, we have $v_1\neq v_2$.
(2) The second case requires that
for any variable $v$ with two concatenations $v=v_1+ v_2$ and $v=v_3+ v_4$,
$\{v_1,v_2\}$ and $\{v_3,v_4\}$ are disjoint.
(3) The third case  requires
that for any two different variables $v$ and $v'$ with 
the concatenation constraints $v=v_1+ v_2$ and $v'=v_3+ v_4$,
$\{v_1,v_2\}$ and $\{v_3,v_4\}$ are disjoint.

With this definition we have the following theorem.

\begin{theorem}[Completeness for the  tree property]
	\label{thm-forward_sound_tree} 
	~\\
	\begin{tabular}{rl}
		\fixes & $S$ $\mathit{Concat}$ $\mathit{Reg}$\\
		\assumes & 
		1. $\mathbf{wf}~S~\mathit{Concat}~\mathit{Reg}$ \\
		& 2. $(\exists l.~$ $\mathbf{acyclic}~S~\mathit{Concat}~l~)$\\
		&3. $\mathbf{tree}~\mathit{Concat}$\\
		\shows &
		$\text{$\mathbf{Forward\_Prop}$}(S, \mathit{Concat}, \mathit{Reg})\leq\mathbf{SPEC}(\lambda\mathit{Reg'}.$\\
		& $((\forall v.~\mathcal{L}(\mathit{Reg}'~v)\neq\emptyset) \longleftrightarrow$ \\
		& $\quad(\exists \mu.~ \mathbf{sat\_str}~S~\mathit{Concat}~\mathit{Reg}~\mu)))
		$\\
	\end{tabular}
\end{theorem}

This theorem shows that
if $\mathit{Concat}$ satisfies the tree property then
we have:
$\mathit{Reg}'$ satisfies that for all $v$, the language of $\mathit{Reg}'~v$ is not empty iff there exists an assignment, which makes the string constraint satisfiable.

In order to check whether a string constraint satisfies the tree property, we propose Algorithm \ref{alg-tree_checking}.
It first stores all the variables, which are on the right hand side of the concatenation constraints, into a list,
and then checks whether the list is distinct.
If it is not distinct then there must exist a variable with at least indegree 2 and thus $\mathit{Concat}$ cannot satisfy the tree property.

\begin{algorithm}
	\caption{Tree property checking}
	\label{alg-tree_checking}
	\begin{algorithmic}[1]
		\Procedure{$\mathbf{Check\_Tree}$}{$S$, $\mathit{Concat}$}~~
		\State $l$ $\gets$ $[]$
		\For{each $v\in S$}
		\For{each $(v_1,v_2)\in (\mathit{Concat}~v)$}
		\State $l\gets v_1~\#~v_2~\#~l$
		\EndFor
		\EndFor
		\State \textbf{return} $\mathbf{distinct}~l$ 
		\EndProcedure
	\end{algorithmic}
\end{algorithm}

Now we show that if Algorithm \ref{alg-tree_checking}
returns $\mathit{True}$ then the tree property is satisfied.

\begin{theorem}[Correctness of $\mathbf{Check\_Tree}$]
	\label{thm-check-tree} 
	~\\
	\begin{tabular}{rl}
		\fixes & $S$ $\mathit{Concat}$\\
		\assumes & 
		1. $dom~\mathit{Concat}~\subseteq S$\\
		&2.  $(\forall v~v_1~v_2. (v_1,v_2)\in \mathit{Concat}~v\longrightarrow \{v_1,v_2\}\subseteq S)$\\
		\shows &  $\mathbf{Check\_Tree}(S,~\mathit{Concat})\longrightarrow \mathbf{tree}~\mathit{Concat}
		$\\
	\end{tabular}
\end{theorem}

With the function $\mathbf{Check\_Tree}$, \tool~can return
three results: \emph{unsat}, \emph{sat}, and \emph{unknown},
in which \emph{unknown} means the string constraint does not satisfy
the tree property and the forward-propagation does not refine the regular constraint of any variable to empty.

We now finish introducing
the forward-propagation procedure.
As it uses many NFA operations,
in the following section, we will
introduce our implementation for these NFA operations.

\section{Symbolic Automata Formalization}
\label{sec-SAF}

Automata operations are the key to our certified string solver. 
For instance, we need to check the language emptiness of 
an NFA as well as construct the concatenation and product of two NFAs.
In order to make the string solver  practically useful, we should also keep
the efficiency in mind.

As argued by D'Antoni et. al \cite{DBLP:conf/cav/DAntoniV17}, the classical definition of NFA, 
whose transition labels  are a character in the alphabet, is not suitable for practical implementation, and they propose s-NFAs, which allow transition labels to carry
a set of characters, to address this limitation.
In this section, we present our formalization of s-NFAs in Isabelle.

\begin{definition}[Abstract s-NFAs]
	\label{def-sym-nfa}
	An s-NFA is a 5-tuple:
	$\mathcal{A}=(Q,\Sigma,\Delta,I,F)$, where $Q$ is a finite set of states,
	$\Sigma$ is an alphabet ($\Sigma$ may be infinite), 
	$\Delta\subseteq Q \times 2^\Sigma\times Q$ is a finite set of transition relations,
	$I\subseteq Q$ is a set of initial states, and
	$F\subseteq Q$ is a set of of accepting states.
\end{definition}

For a transition $(q,\alpha,q')$ in an  s-NFA, the implementation of $\alpha$
can use various different data structures, such as intervals for string solvers and 
Binary Decision Diagrams (BDD) for symbolic model checkers.
Definition \ref{def-sym-nfa}
of s-NFAs is called abstract s-NFAs because the transition labels
are denoted as sets instead of using concrete data structures, like intervals and 
BDDs.



In order to make our formalization of s-NFAs reusable for different implementation 
of transition labels, we exploit Isabelle's refinement framework and 
divide the formalization of s-NFAs in two levels:
(1)~the abstract level (introduced in this section) and 
(2)~the implementation level (introduced in the following section).
When the correctness of the algorithms at the abstract level  is proven,
the implementation level does not need to re-prove it, only the refinement relations between these two levels
must be shown.

At the abstract level, transition labels are modeled as a set as shown in Definition \ref{def-sym-nfa}.
At the implementation level, the sets of characters are refined to concrete data structures, such as intervals or BDDs.
\begin{figure}
{
   \begin{lstlisting}
record NFA = Q :: "'q set"
             |$\Delta$| :: "('q*'a set*'q) set"
             I :: "'q set"
             F :: "'q set"
        
definition |$\mathbf{well}\text{-}\mathbf{formed}$| where
|$\mathbf{well}\text{-}\mathbf{formed}~\mathcal{A} = (I~\mathcal{A}) \subseteq (Q~\mathcal{A})\land (F~\mathcal{A})\subseteq (Q~\mathcal{A})$|
              |$\land~\mathbf{finite}$ $(Q$ $\mathcal{A}) \land \mathbf{finite}$ $\Delta$ $\land$|
          |$(\forall (q,\alpha,q')\in \Delta, q\in (Q$ $\mathcal{A}) \land q'\in (Q$ $\mathcal{A}))$|
               
fun |$\mathbf{reachable}$| w |$\Delta$| q q' where
  |$\mathbf{reachable}$| [] |$\Delta$| q q' = (q=q')
 |$\mid\mathbf{reachable}$| (a |$\#$| w) |$\Delta$| q q' = 
          (|$\exists q_i~\alpha.$| (q,|$\alpha$|,|$q_i$|)|$\in$||$\Delta$|
          |$~\land~$|a|$\in\alpha\land\mathbf{reachable}$| w |$\Delta$| |$q_i$| q')

definition |$\mathbf{NFA\_accept}$| where
|$\mathbf{NFA\_accept}$| w |$\mathcal{A} = (\exists q$ $q'.$ $q \in ($I $\mathcal{A}) \land q' \in ($F $\mathcal{A})$| 
                |$\land~\mathbf{reachable}$ w $(\Delta$ $\mathcal{A})$ q q'$)$|
             
definition |$\mathcal{L}$| where
|$\mathcal{L}$| |$\mathcal{A} =$| |$\{w.$| |$\mathbf{accept}$| w |$\mathcal{A}\}$|
         
   \end{lstlisting}
}
\caption{Isabelle formalization of abstract symbolic NFA}
\label{label-isabelle-NFA}
\end{figure}

Our formalization of s-NFAs extends
an existing classical NFA formalization developed by Tuerk et al.~\cite{Tuerk-NFA}; this formalization is part of the verified CAVA model checker~\cite{Julian}.
We extend it to support s-NFAs and further operations, such as the concatenation operation
of two NFAs.
Figure \ref{label-isabelle-NFA} shows the Isabelle formalization of s-NFAs.
An s-NFA is defined by a record with four elements:
(1)~$\texttt{Q}$ is the set of states, 
(2)~$\texttt{I}$ is the set of initial states, 
(3)~$\texttt{F}$ is the  set of accepting states, and
(4)~$\Delta$ is the set of transitions with transition labels defined by
     sets.
The definition $\mathbf{well}$-$\mathbf{formed}$ is a predicate to check
whether an NFA satisfies (1)~the states occurring
in the sets of initial and accepting states are also in the set (\texttt{Q $\mathcal{A}$}) of states.
In Isabelle,
the value of a field $\mathtt{fd}$ in a record $\mathtt{rd}$ can be extracted by $\mathtt{(fd~rd)}$.
Therefore,  (\texttt{Q $\mathcal{A}$}) 
denotes the value of the field \texttt{Q} in $\mathcal{A}$. 
(2)~The states in the set of transitions should also be in (\texttt{Q $\mathcal{A}$}).
(3)~(\texttt{Q $\mathcal{A}$}) is a finite set.
(4)~$\Delta$ is finite.

For a word $\texttt{w}=a_1a_2\ldots a_n$ and two states \texttt{q} and \texttt{q'},
the function $\mathbf{reachable}$ checks whether 
there exists a path $q_0, \alpha_1,$ $q_1, \ldots, q_{n-1}, \alpha_n, q_n$ in the NFA $\mathcal{A}$ such that $\texttt{q} = q_0\land \texttt{q'}=q_n$,
$(q_{i-1},\alpha_i,q_{i}) \in (\Delta~\mathcal{A}), 1 \leq i \leq n$ and
$a_j\in \alpha_j, 1\leq j \leq n$.
The definition $\mathbf{NFA\_accept}$
checks whether a word is accepted by an NFA 
and the definition of $\mathcal{L}$ is the language of an NFA.

Based on this definition of NFAs, a collection of automata operations are defined and their correctness lemmas are proven.
Here we only present the operation of the concatenation shown in Figure \ref{fig-def-concatenation}.

\begin{figure}
		\begin{lstlisting}
definition |$\mathbf{NFA\_concat\_basic}$| where 
|$\mathbf{NFA\_concat\_basic}$| |$\mathcal{A}_1$| |$\mathcal{A}_2 \equiv \llparenthesis$|
  Q = (Q |$\mathcal{A}_1$|) |$\cup$| (Q |$\mathcal{A}_2$|), 
  |$\Delta$| = |$(\Delta$ $\mathcal{A}_1) \cup(\Delta$ $\mathcal{A}_2)$ $\cup$|
      |$\{(q,\alpha,q'')\mid \exists q'. (q,\alpha, q')\in (\Delta$ $\mathcal{A}_1)\land$|
               |$q'\in (F$ $\mathcal{A}_1)\land q''\in ($I $\mathcal{A}_2)\}$|,
 I = if ((I |$\mathcal{A}_1)$ $\cap$ |(F |$\mathcal{A}_1)=\emptyset$|) then (I |$\mathcal{A}_1$|)
       else ((I |$\mathcal{A}_1$|) |$\cup$| (I |$\mathcal{A}_2$|)), 
 F = (F |$\mathcal{A}_2$|)
|$\rrparenthesis$|

definition |$\mathbf{NFA\_concat}$| where
|$\mathbf{NFA\_concat}$| |$\mathcal{A}_1$| |$\mathcal{A}_2$ $f_1$ $f_2 =$|
 |$\mathbf{remove\_unreachable\_states}$|
   (|$\mathbf{NFA\_concat\_basic}$|
     (|$\mathbf{NFA\_rename}$| |$f_1$| |$\mathcal{A}_1$|)(|$\mathbf{NFA\_rename}$| |$f_2$| |$\mathcal{A}_2$|))
		\end{lstlisting}
	\caption{The definition of concatenation operation}
	\label{fig-def-concatenation}
\end{figure}

$\mathbf{NFA\_concat\_basic}$ constructs the concatenation of two NFAs. 
The correctness of this definition relies on the fact that the sets of states in the two NFAs are disjoint.
In the definition of $\mathbf{NFA\_concat\_basic}$,
the set of transitions of the concatenation
is the union of (1) $(\Delta~\mathcal{A}_1)$,  (2) $(\Delta~\mathcal{A}_2)$, and
(3) $\{(q,\alpha,q'')\mid \exists q'. (q,\alpha, q')\in (\Delta~\mathcal{A}_1)\land q'\in (F~\mathcal{A}_1)\land q''\in (I~\mathcal{A}_2)\}$.
The transitions in (3) concatenate
the language of $\mathcal{A}_1$ to the language of $\mathcal{A}_2$.

The set of the initial states in the concatenation
is computed by 
first checking whether there is a state that is in both the sets of initial and accepting states
of $\mathcal{A}_1$.
If there exists such a state then
the initial states in the concatenation
should include the initial states of $\mathcal{A}_2$,
as $\mathcal{A}_1$ accepts the empty word,
otherwise the initial states of the concatenation are
$\mathcal{A}_1$'s initial states.

The definition of $\mathbf{NFA\_concat}$ 
firstly renames the states in both NFAs 
to ensure their sets of states disjoint.
In the term ``$\mathbf{NFA\_rename}$ $f$ $\mathcal{A}$'',
$f$ is a renaming function, for instance, 
$f$ can be ``$\lambda q. (q,1)$'', which renames
 any state $q$ to $(q,1)$. 
``$\mathbf{NFA\_rename}~f$ $\mathcal{A}$''
renames all states in $($Q $\mathcal{A})$
by the function $f$.
Correspondingly, all states in transitions, initial states, and accepting states
are also renamed by $f$.
Renaming states makes it easier to ensure the sets of states of two NFAs are disjoint.
The function $\mathbf{remove\_unreachable\_ states}$
removes all states in an NFA that are not reachable from the initial states of the NFA.
This enables us to check the language emptiness of an NFA by only checking
the emptiness of its set of accepting states.
At the implementation level, the algorithm also only  generates reachable states for NFAs.
But with $\mathbf{remove\_unreachable\_ states}$ used in
$\mathbf{NFA\_concat}$ at the abstract level,
the refinement relation between these two levels
can be specified as the NFA  \emph{isomorphism} of the two output concatenations of automata at the two levels, instead of language equivalence, which is more challenge to prove.

In order to ensure the correctness of this operation,  we need to prove the 
following lemma in Isabelle.

\begin{lemma}[Correctness of $\mathbf{NFA\_concat}$]
	\label{lem-correct_concatenation} 
	~\\
	\begin{tabular}{rl}
		\fixes & $\mathcal{A}_1$ $\mathcal{A}_2$ $f_1$ $f_2$\\
		\assumes 
			& 1. $\mathbf{well}$-$\mathbf{formed}$ $\mathcal{A}_1$\\
			& 2. $\mathbf{well}$-$\mathbf{formed}$ $\mathcal{A}_2$\\
			& 3. $(\mathbf{image}$ $f_1$ $($\texttt{Q} $\mathcal{A}_1))$ $\cap$ $(\mathbf{image}$ $f_2$ $($\texttt{Q} $\mathcal{A}_2))=\emptyset$\\
			& 4. $\mathbf{inj\_on}$ $f_1$ $($\texttt{Q} $\mathcal{A}_1)\land$ $\mathbf{inj\_on}$ $f_2$ (\texttt{Q} $\mathcal{A}_2)$\\
		\shows &  $\mathcal{L}(\mathbf{NFA\_concat}~\mathcal{A}_1~\mathcal{A}_2~f_1~f_2)=$\\
		& $\{w_1@w_2.~w_1 \in \mathcal{L}(\mathcal{A}_1)\land w_2\in \mathcal{L}(\mathcal{A}_2)\}$\\
	\end{tabular}
\end{lemma}
The assumptions $\mathbf{well}$-$\mathbf{formed}$$~\mathcal{A}_1$ and $\mathbf{well}$-$\mathbf{formed}$ $\mathcal{A}_2$
require the input NFAs to be well-formed. 
In addition, 
the assumption ($\mathbf{image}$ $f_1$ $($\texttt{Q} $\mathcal{A}_1)$) $\cap$ ($\mathbf{image}$ $f_2$ $($\texttt{Q} $\mathcal{A}_2)$)$=\emptyset$ requires that
after renaming the states in both $\mathcal{A}_1$ and $\mathcal{A}_2$,
the new sets of states of the two NFAs are disjoint.
The term ``$\mathbf{image}$ $f$ $S$'' applies the function $f$
to the elements in the set $S$ and returns a new set $\{f e.~e\in S\}$.
The premises
$\mathbf{inj\_on}~f_1~(\texttt{Q}~\mathcal{A}_1)$ and
$\mathbf{inj\_on}~f_2~(\texttt{Q}~\mathcal{A}_2) $ require that
the functions $f_1$ and $f_2$ are injective over the sets $(\texttt{Q}~\mathcal{A}_1)$
and $(\texttt{Q}~\mathcal{A}_2)$, respectively.
The conclusion in ``\shows'' specifies that
the language of the concatenation of  $\mathcal{A}_1$ and $\mathcal{A}_2$
equals the set of the concatenations of the words in  $\mathcal{L}(\mathcal{A}_1)$ and the words in $\mathcal{L}(\mathcal{A}_2)$. 


Moreover, some other important functions are defined, for instance:
\begin{itemize}


\item The product of two NFAs: $\mathbf{NFA\_product}~\mathcal{A}_1~\mathcal{A}_2$, which computes the product 
of $\mathcal{A}_1$ and $\mathcal{A}_2$. We proved the theorem:  $\mathcal{L}(\mathbf{NFA\_product}~\mathcal{A}_1~\mathcal{A}_2) = \mathcal{L}(\mathcal{A}_1)\cap\mathcal{L}(\mathcal{A}_2)$.

\item Checking the isomorphism of two NFAs: \\
$\mathbf{NFA\_isomorphism}$ $\mathcal{A}_1~\mathcal{A}_2\triangleq(\exists f. \mathbf{inj\_on}~f~(\texttt{Q}~\mathcal{A}_1)~\land~$
$\mathbf{NFA\_rename}~f~\mathcal{A}_1=\mathcal{A}_2)$.
Some lemmas are proven for it, such as,
$\mathbf{well}$-$\mathbf{formed}$ $\mathcal{A}_1\land$
$\mathbf{well}$-$\mathbf{formed}$ $~\mathcal{A}_2\land\mathbf{NFA\_isomorphism}~\mathcal{A}_1~\mathcal{A}_2 \longrightarrow \mathcal{L}(\mathcal{A}_1) = \mathcal{L}(\mathcal{A}_2)$.
\end{itemize}

 Section \ref{sec-forward-analysis}  and this section present the abstract level algorithms for the forward-propagation and the
NFA operations, respectively.
In order to make \tool~efficient and practically useful, we 
need to use some efficiently implemented data structures, such as red-black-trees (RBT) and 
hash maps.
This will be introduced in the following section.

\section{Implementation-Level Algorithms}

In this section,
we will first present the relations between abstract concepts, like sets and maps, 
and the implementation data structures in Isabelle (Subsection \ref{sec-icf-new}),
which is the prerequisite to understand
the implementation-level algorithm
in Subsection \ref{sec-implementation-new}.

\subsection{Isabelle Collections Framework}
\label{sec-icf-new}
Isabelle Collections Framework (ICF) \cite{DBLP:conf/itp/LammichL10} provides an efficient, extensible, and machine checked collections framework. 
The framework features the use of data refinement techniques  \cite{DBLP:conf/itp/Lammich13} to refine an abstract specification (using high-level concepts like sets) to a more concrete implementation (using collection data structures, like RBT and hashmaps). 
The code-generator of Isabelle can be used to generate efficient code.

The concrete data structures implement a collection of interfaces that mimic the operations of
abstract concepts.
We list the following RBT implementation of set interfaces that will be used to present the implementation-level algorithms:
\begin{itemize}
	\item $RBT.\mathbf{empty}$, generates an RBT representation of the empty set.
	\item $RBT.\mathbf{mem}~e~D$, checks whether the element $e$ is in the set represented by  RBT $D$.
	\item $RBT.\mathbf{insert}~e~D$, inserts the element $e$ into the set represented by RBT $D$.
	\item $RBT.\mathbf{inter}~D_1~D_2$, computes the RBT representation of the set intersection of $D_1$ and $D_2$. $D_1$ and $D_2$ are both RBT represented sets.
	\item $RBT.\mathbf{union}~D_1~D_2$, computes the RBT representation of the set union of $D_1$ and $D_2$. $D_1$ and $D_2$ are both RBT represented sets.
	\item $RBT.\mathbf{iterator}~D~v~f$, where $f$ is of the type $t_1\Rightarrow t_2 \Rightarrow t_2$,  $v$ is of the type $t_2$, and the elements in $D$ are of the type $t_1$. This interface  implementation iteratively applies the function $f$ to the elements in $D$ with the initial value $v$.
	For instance, the first iteration of the computation of ``$RBT.\mathbf{iterator}~\{e_1,e_2\}~v~f$''
	 selects an element from $\{e_1,e_2\}$, assume $e_1$ is selected,
	and then 
	 applies $f$ to $e_1$ and $v$, i.e., ``$f~e_1~v$''.
	Assume the value of ``$f~e_1~v$'' is $v'$.
	The second iteration applies  $f$ to $e_2$ and $v'$.
	Now all elements in the set are traversed, ``$RBT.\mathbf{iterator}~\{e_1,e_2\}~v~f$''
	finishes with the return value of ``$f~e_2~v'$''.
	\item $RBT.\mathbf{to\_list}~D$ translates a set represented by $D$ to a list.
	\item $RBT.\alpha~D$ translates RBT represented set $D$ to the set in the abstract concept.
\end{itemize}

%

\subsection{Implementation-Level Algorithm for $\mathbf{NFA\_concat}$ }
\label{sec-implementation-new}

In this subsection, we present the implementation-level algorithm by the NFA operation 
$\mathbf{NFA\_concat}$.

In order to formalize implementation-level algorithms,
the first step is to consider the data refinement.
At the abstract level, we use sets to store states, transitions, initial states,
and accepting states.
At the implementation level, we use RBTs to replace sets to store these elements of NFAs.
The type of NFAs using RBT data structures is specified in Figure \ref{NFA-RBT-code}. {Note that, here we only give a simplified version of the RBT NFA definition to make it easy to understand. Indeed, RBT has more type arguments}.

Moreover, for each transition $(q,\alpha,q')$, the label $\alpha$
is represented by a set at the abstract level.
We use intervals to replace sets to store $\alpha$.
An interval $[n_1,n_2]$ is a pair of elements in a totally ordered set.
The semantics of an interval 
is defined by a set as $\mathbf{semI}~[n_1,n_2]=\{n\mid n_1\leq n\leq n_2\}$.

\begin{figure}
	{
		\begin{lstlisting}
record NFA_rbt = 
      Q :: "'q RBT"
      |$\Delta$| :: "('q * 'a Interval * 'q) RBT"
      I :: "'q RBT"
      F :: "'q RBT"
		\end{lstlisting}
	}
	\caption{NFA data structure definition using RBT}
	\label{NFA-RBT-code}
\end{figure}

In order to support NFA operation implementation,
some operations for intervals are implemented.
We list the following $3$ interval operations here.
\begin{itemize}
	\item Computing the intersection of two intervals: \textbf{intersectionI} $[I_1,I_2]$ $[I_1',~I_2'] \triangleq [\mathbf{max}(I_1,I_1'), \mathbf{min}(I_2,I_2')]$. The term $\mathbf{max}(I_1,I_1')$ returns the bigger one of $I_1$ and $I_1'$, and
    $\mathbf{min}(I_2,I_2')$ returns the smaller one of $I_2$ and $I_2'$.
	\item Checking the non-emptiness of the interval $[I_1,I_2]$: \textbf{nemptyI} $[I_1,I_2] \triangleq I_1 \leq I_2$.
	\item Checking the membership of an element $e$ and the interval $[l_1,l_2]$: \textbf{memI} $e$ $[l_1,l_2]\triangleq l_1 \leq e \leq l_2$. 
\end{itemize}

\begin{algorithm}
	\caption{The algorithm idea of NFA concatenation}
	\label{alg-concate}
	\begin{algorithmic}[1]
		\Procedure{{$\mathbf{Construct\_Trans}$}}{$S_{\Delta1},S_{\Delta2}, S_{I2}, S_{F1}$}
		\State $S_D$ $\gets$ $RBT.\mathbf{union}~S_{\Delta1}~S_{\Delta2}$;
		\State $RBT.\mathbf{iterator}$ $S_{\Delta1}$ $S_D$
		\State \quad$(\lambda (q,a,q'')~S.$ ~~
		\State \quad \quad$\mathbf{if}~ (RBT.\mathbf{mem}~q''~S_{F1})~\mathbf{then}~$
		\State \quad\quad $~~RBT.\mathbf{iterator}$ $S_{I2}$ $S$
		\State \quad\quad\quad $(\lambda q'~S'.~ RBT.\mathbf{insert}~(q,a,q')~S'$) 
		\State \quad\quad \textbf{end~if});
		\State \textbf{return} $S_D$;
		\EndProcedure\\
		
		\Procedure{{$\mathbf{NFA\_Concate\_Impl}$}}{$\mathcal{A}_1$, $\mathcal{A}_2$, $f_1$, $f_2$}
		\State \Comment{$\mathcal{A}_1$, $\mathcal{A}_2$ are of type \texttt{NFA\_rbt}}
		\State $\mathcal{A}_1'\gets\mathbf{NFA\_Rename\_Impl}~f_1~\mathcal{A}_1$;
	    \State $\mathcal{A}_2'\gets\mathbf{NFA\_Rename\_Impl}~f_2~\mathcal{A}_2$
;		\If {$RBT.\mathbf{inter}~(\texttt{I}~\mathcal{A}_1')~(\texttt{F}~\mathcal{A}_1')\neq RBT.\mathbf{empty}$} \label{initial-begin}
		\State $S_I\gets RBT.\textbf{union}~(\texttt{I}~\mathcal{A}_1')~(\texttt{I}~\mathcal{A}_2')$
		\Else 
		\State $S_I\gets (\texttt{I}~\mathcal{A}_1')$
		\EndIf; \label{initial-end}
		\State $S_Q\gets S_I$; $S_\Delta\gets RBT.\mathbf{empty}$;  $wl\gets RBT.\mathbf{to\_list}~S_Q$; \label{qd_initial}
		\State $S_\Delta'\gets \mathbf{Construct\_Trans}(\Delta~\mathcal{A}_1',\Delta~\mathcal{A}_2', \texttt{I}~\mathcal{A}_2', \texttt{F}~\mathcal{A}_1')$;
		\While{$wl\neq []$}\Comment{``$[]$'' denotes empty list}\label{while-begin}
		\State $q_s\gets wl.\mathbf{rm\_first}$ ;
		\State \Comment{``$\mathbf{rm\_first}$'': removes and returns the first} 
		\State \quad \quad \quad \quad \quad \quad \quad \quad element in $wl$
		\State $RBT.\mathbf{iterator}$ $S_\Delta'$
		$S_\Delta$
		\State \quad$(\lambda(q,\alpha,q')~S.$
		\State \quad\quad$\mathbf{if}~q=q_s\land (\mathbf{nemptyI}~\alpha)$ \textbf{then}
		\State \quad\quad\quad$\mathbf{if}~\neg(RBT.\mathbf{mem}~q'~S_Q)$ \textbf{then}
		\State \quad\quad\quad\quad$RBT.\mathbf{insert}~q'~S_Q$;
		\State \quad\quad\quad\quad$wl\gets wl@[q']$; 
		\State \quad\quad\quad\textbf{end~if};
		\State \quad\quad\quad $RBT.\mathbf{insert}~(q,\alpha,q')~S$;
		\State \quad\quad \textbf{end~if};
		\State \quad$)$
		\EndWhile \label{while-end}
		\State $\mathbf{return}~(S_Q,S_\Delta,S_I,RBT.\mathbf{inter}~S_Q~(F~\mathcal{A}_2'))$ \label{return-label}
		\EndProcedure
	\end{algorithmic}
\end{algorithm}

Algorithm \ref{alg-concate} shows the basic idea of the $\mathbf{NFA\_concat}$ implementation (the procedure $\mathbf{NFA\_Concat\_Impl}$).
The texts after ``$\triangleright$'' are comments.
$\mathbf{NFA\_Rename\_Impl}$ is the implementation
of the NFA renaming function $\mathbf{NFA\_Rename}$ at the abstract level.

The sub-procedure $\mathbf{Construct\_Trans}$ generates
the set of transitions for the concatenation of the two automata,
which contains the transitions in both $({\Delta}~\mathcal{A}_1)$ and $({\Delta}~\mathcal{A}_2)$, and
the transitions that concatenate the two automata (cf. Figure \ref{fig-def-concatenation}
for the abstract algorithm of the concatenation ).

In Algorithm \ref{alg-concate},
Line \ref{initial-begin}-\ref{initial-end} computes the initial states of the concatenation.
Line \ref{qd_initial} initializes
$S_Q$, $S_\Delta$, and $wl$,
where $S_Q$ stores the states of the concatenation NFA,
$S_\Delta$ stores the transitions of the concatenation NFA,
$wl$ is a list to mimic a queue that stores the
states to be expanded.
Line \ref{while-begin} to \ref{while-end} 
is the while loop for expanding the reachable states and transitions.
Finally Line \ref{return-label} returns the concatenation of the two NFAs.
This algorithm constructs the concatenation of two NFAs 
with only reachable states, which means that 
we can check the emptiness of the concatenation NFA by only checking
the emptiness of its set of accepting states.

In order to ensure the correctness of the implementation for concatenation, 
we need to prove the refinement relation between $\mathbf{NFA\_concat}$ and $\mathbf{NFA\_Concate\_Impl}$.

In Figure \ref{NFA-refine-code}, we formalize the refinement relations between the two 
NFAs at the abstract level and the implementation level.
The function $\mathbf{NFA}$\_$\alpha$ translates
an implementation-level NFA $\mathcal{A}$ (its type is \texttt{NFA\_rbt})
to an abstract-level NFA of the type \texttt{NFA}.
The definition $\mathbf{refine\_rel}$ defines the 
refinement relation between an abstract-level NFA $\mathcal{A}$
and an implementation NFA $\mathcal{A}'$, which requires
the isomorphism between $\mathcal{A}$ and  $(\mathbf{NFA}\_\alpha~\mathcal{A}')$

Lemma \ref{lemma_correct_concat_impl} specifies that the concatenation implementation correctly
refines the abstract concatenation definition.

\begin{figure}
	{
		\begin{lstlisting}
definition |$\mathbf{NFA}$|_|$\alpha$| where
|$\mathbf{NFA}$|_|$\alpha$| |$\mathcal{A}$| |$\equiv$| |$\llparenthesis$| Q = |$RBT.\alpha$| (Q |$\mathcal{A}$|), 
           |$\Delta$| = (|$\mathbf{image}$| |$(\lambda(q,\alpha,q').(q,\mathbf{semI}~\alpha,q'))$| 
                        (|$RBT.\alpha$| (|$\Delta$| |$\mathcal{A}$|))), 
           I = |$RBT.\alpha$| (I |$\mathcal{A}$|), 
           F = |$RBT.\alpha$| (F |$\mathcal{A}$|)|$\rrparenthesis$|	
           
definition |$\mathbf{refine\_rel}$| where
|$\mathbf{refine\_rel}$| |$\mathcal{A}$| |$\mathcal{A}'$|| $=$|
        |$\mathbf{NFA\_isomorphism}$| (|$\mathbf{NFA}$|_|$\alpha$| |$\mathcal{A}'$|) |$\mathcal{A}$|
             
lemma refine_rel_lang: 
    |$\mathbf{well}$|-|$\mathbf{formed}$ $\mathcal{A}$| |$\land$| |$\mathbf{refine\_rel}$| |$\mathcal{A}$| |$\mathcal{A}'$|
          |$\Longrightarrow$| |$\mathcal{L}(\mathcal{A})$| = |$\mathcal{L}($$\mathbf{NFA}$\_$\alpha$ $\mathcal{A}')$|
		\end{lstlisting}
	}
	\caption{Refinement relation formalization}
	\label{NFA-refine-code}
\end{figure}

\begin{lemma}[Correctness of $\mathbf{NFA\_Concate\_Impl}$]\label{lemma_correct_concat_impl}~\\
	\fixes~$\mathcal{A}_1$, $\mathcal{A}_2$, $\mathcal{A}_1'$, $\mathcal{A}_2'$, $f_1$, $f_2$, $f_3$, $f_4$\\
	\assumes
	\begin{enumerate}		
		\item $\mathbf{refine\_rel}$ $ \mathcal{A}_1~\mathcal{A}_1'~\land~\mathbf{refine\_rel}$ $ \mathcal{A}_2~\mathcal{A}_2'$
		\item  $\mathbf{inj\_on}$ $f_1$ $($\texttt{Q} $\mathcal{A}_1)\land\mathbf{inj\_on}$ $f_2$ $($\texttt{Q} $\mathcal{A}_2)$
		\item  $(\mathbf{image}$ $f_1$ $($\texttt{Q} $\mathcal{A}_1))$ $\cap$ $(\mathbf{image}$ $f_2$ $($\texttt{Q} $\mathcal{A}_2))=\emptyset$
		\item  $\mathbf{inj\_on}$ $f_3$ $(RBT.\alpha~($\texttt{Q} $\mathcal{A}_1'))\land\mathbf{inj\_on}$ $f_4$ $(RBT.\alpha~($\texttt{Q} $\mathcal{A}_2'))$
		\item  $($$\mathbf{image}$ $f_3$ $(RBT.\alpha$ $($\texttt{Q} $\mathcal{A}_1')))$ $\cap$ $($$\mathbf{image}$ $f_4$ $(RBT.\alpha$~$($\texttt{Q} $\mathcal{A}_2')))$$=\emptyset$
		\item $\mathcal{A}=\mathbf{NFA\_concat}(\mathcal{A}_1,\mathcal{A}_2,f_1,f_2)$
		\item 		   $\mathcal{A}'=\mathbf{NFA\_Concate\_Impl}(\mathcal{A}_1',\mathcal{A}_2',f_3,f_4) $
	\end{enumerate}
	\shows~
	$\mathbf{refine\_rel}$ $\mathcal{A}$ $\mathcal{A}'$
\end{lemma}

This theorem shows that the implementation-level 
concatenation of two NFAs is a refinement for the concatenation
of the corresponding  abstract-level NFAs.
With Lemma \ref{lemma_correct_concat_impl} and 
Lemma \texttt{refine\_rel\_lang}, we can easily infer that 
the language of the concatenation at the abstract level
is equal to the language of the concatenation at the implementation level.

In this section, we presented the implementation of the concatenation operation.
This implementation uses intervals to store transition labels.
Indeed, our formalization of s-NFAs can be reused 
to implement s-NFAs using other data structures, such as BDDs, bitvectors, and Boolean formulas, to store transition labels.
The corresponding operations, such as \textbf{intersectionI},
\textbf{nemptyI}, and \textbf{memI}, for these data structures  need to be implemented at the implementation level accordingly.

Moreover, the forward-propagation algorithm and other automata operations  
also have their corresponding implementations
and their refinement relations with the abstract algorithms have been proven.

\section{Evaluation and Development Efforts}

The tool \tool~(\url{https://github.com/uuverifiers/ostrich/tree/CertiStr}) is developed in the Isabelle proof assistant (version 2020) and one can extract executable OCaml code from the formalization in Isabelle. To obtain a complete solver, we added a (non-certified) front-end for parsing
SMT-LIB problems (Figure~\ref{fig-workflow}); this front-end is written in Scala, and mostly borrowed
from OSTRICH~\cite{CHL+19}.
In this section, we evaluate \tool~against the Kaluza benchmark~\cite{DBLP:conf/sp/SaxenaAHMMS10}, the most commonly used benchmark to compare string solvers.
Moreover, we also elaborate 
the development efforts for \tool.

\subsection{The Effectiveness and Efficiency}
\label{subsec-effective-efficiency}

The Kaluza benchmark contains 47284 tests,
among which 38043 tests (80.4\%) are in the string constraint
fragment of \tool.
\tool~supports string constraints with word equations of concatenation, regular constraints, and monadic length.
The benchmark
 is classified into $4$ groups according to the results of the Kaluza string solver: (1) sat/small, (2) sat/big, (3) unsat/small, and (4) unsat/big.

 Note that the benchmark classification is not in all cases accurate,
 in the sense that there are some files in unsat/big and unsat/small
 that are satisfiable, owing to known incorrect results that were
produced by the original Kaluza string solver~\cite{DBLP:conf/sp/SaxenaAHMMS10,cvc4}.

\begin{table*}[tp]
	\vspace{2mm}
	\centering
	\caption{Experimental results of the Kaluza benchmark}
	\label{tab-experimental-results}
	\begin{tabular}[t]{l||ccccccc|c}
		\hline
		&~~total tests~~&~~sat~~&~~unknown~~&~~unsat~~&~~solved\%~~&~~avg. time(s)~~&~~timeout~~&~~CVC4~~\\
		\hline
		\hline
		~~~sat\_small & 19634 & 19302 & 332 & 0 & 98.3\% & < 0.01& 0 & < 0.01 \\
		~~~sat\_big & 774 & 521 & 253 & 0 & 67.3\%& 0.84 & 0 & 0.04\\
		~~~unsat\_small~~~&8775&7376 & 824 & 575 & 91\% & 0.58&  0& < 0.01\\
		~~~unsat\_big&8860& 2502 &4880 & 1478 & 45\% & 8.01&  728 & 0.5\\
	\hline
		~~~total & 38043 & 29700 & 6289 &2054 & 83.5\% & 1.87 & 728 & 0.11\\ 
		\hline
	\end{tabular}
\end{table*}%

\begin{table*}[t]
	\centering
	\caption{Experimental results for state and transition explosion}
	\label{table-time}
	\begin{tabular}[t]{r|c|c|c|c|c|c}
		\hline
		\emph{No. Concat} & 11 & 12 & 13 & 14 & 15 & 16\\
		\hline
		\emph{states}~~~& ~~2048~~ & ~~4096~~ & ~~8192~~ & ~~16384~~ & ~~32768~~  & ~~65536~~ \\
		\emph{transitions}~~~& 177147 & 531441 & 1594323 & 4782969 & 14348907 & 43046721\\
		\hline\hline
		\emph{time} (s)~~~& 0.43 & 1.29 & 4.10 & 12.37 &  50.59  & 168.28\\
		\hline
	\end{tabular}
\end{table*}%

We ran \tool~over
these 38043 tests on an AMD Opteron 2220 SE machine, running 64-bit Linux and Java~1.8 (for running the Scala front-end of \tool).
The results are shown in Table \ref{tab-experimental-results}.
The column ``total tests'' denotes
the total number of tests in a group.
The columns ``sat'', ``unknown'' , ``unsat'' denote
the numbers of tests for which \tool~returns \emph{sat}, 
\emph{unknown}, and \emph{unsat}, respectively.
The column "solved\%" denotes
the percentage of the tests for which the string solver returns \emph{sat} or \emph{unsat}.
The columns "avg.time(s)" and ``timeout" denote the average time 
for running each test in \tool~and the number of tests that time out, respectively.
The column "CVC4" denotes the average execution time of the state of the art string solver CVC4, which 
can efficiently solve all the string constraints without timeout.
The time limit for solving each test is 60 seconds.

From Table \ref{tab-experimental-results}, we can conclude that 
in total, 83.5\% of the tests can be solved by \tool~
with the result \emph{sat} or \emph{unsat}.
The remaining of the tests are unknown or timeout.
The groups sat\_small and unsat\_small   
have the higher solved percentages, more than 90\%, compared with
sat\_big and unsat\_big.
This is easy to understand as small tests
have a higher probability to satisfy the tree property.
The worst group is unsat\_big. \tool~ can
solve 45\% tests in this group.
The reason is that
in this group, there are a lot of tests with more than 
100 concatenation constraints, which significantly increases
their probability of violating the tree property.
For the groups unsat\_big and unsat\_small,
\tool~detects 2502 and 7376 tests, respectively,
that are indeed satisfiable.

Moreover, in unsat\_big, there are 728 tests that time out.
After analyzing these tests, we found that
a variable with a lot of concatenation constraints (i.e.,
the variable appears many times on the left-hand sides of the concatenation constraints)
can easily yield s-NFA state and transition explosion during the forward-propagation.
In order to evaluate such a state explosion problem, we 
set a test 
of the  form in Example \ref{example-stateexposition}: 
\begin{example}
\label{example-stateexposition}
\[
\texttt{x} = \texttt{x}_1~+\texttt{x}_1\wedge~~\texttt{x} = \texttt{x}_2~+\texttt{x}_2\wedge ~~ \texttt{x} = \texttt{x}_3+ \texttt{x}_3\wedge  \ldots
\]
\end{example}

It is a test case with only the variable $x$ on the left-hand side.
On the right-hand side, there are concatenations
of the form $\texttt{x}_i+\texttt{x}_i$.
No regular constraints are in the test.

We ran \tool~over the test case by increasingly adding more concatenations for 
the variable \texttt{x}.
Table \ref{table-time} shows the 
results for solving the test case. 
Each column contains
(1) the number of concatenation constraints (\emph{No. Concat}),
(2) the size of the  automaton of the variable \texttt{x} after 
      executing the forward-propagation (The numbers of \emph{states} and \emph{transitions} of the automaton), and
 (3) the execution time for running the forward-propagation over the test (\emph{time}).

From the table we can conclude that 
after adding $16$ concatenation constraints
for the variable \texttt{x},
the  automaton generated for
\texttt{x} contains
65536 states and 43046721 transitions.
Solving this test takes 168.28s.

Now consider the  728 tests that time out.
These tests have similar constraints, in which some variables have a lot of concatenation constraints.
During the forward-propagation,
state and transition explosion happens for these tests, therefore they cannot be solved in 60s.

Compared with CVC4, which is not automata-based and incorporates more optimizations over  its string theory decision procedure, \tool~is less efficient.
We can also optimize \tool~further in the future.
For instance, we can optimize the language intersection operation $\Sigma^* \cap \mathcal{L}(\mathcal{A})$ to $\mathcal{L}(\mathcal{A})$,
and the language concatenation
$\Sigma^*+\Sigma^*$ to $\Sigma^*$.
We can also implement NFA minimization algorithms to further improve its efficiency. 
These optimizations can avoid the state explosion problem to some extent.
However, integrating such optimizations in a verified solver is challenging,
as more cases need to be proven correct.

\subsection{The Efforts of Developing \tool}

In this subsection, we discuss the efforts of developing \tool.
Table \ref{tab-development-effort} shows the efforts.
The rows ``Abs Automata Lib'' and ``Imp Automata Lib''
denote abstract level and implementation level automata libraries, respectively.
The rows "Abs Forward Prop" and ``Imp Forward Prop''
denote abstract level and implementation level forward-propagations, respectively.
The column ``Loc'' denotes the lines of Isabelle code for each module.
The column ``Terms'' denotes the number of definitions, functions, locales, and classes.
The column ``Theorems'' denotes the number of theorems and lemmas in a module.
The development needs around one person-year efforts.

\begin{table}[h]
	\caption{Development effort of certified string solver}
	\label{tab-development-effort}
\begin{tabular}{c||c|c|c}
	\hline
	Module & Loc & 
	Terms &  Theorems  \\
	\hline\hline
	Abs Automata Lib & 4484 & 100 & 254\\
	Imp Automata Lib & 8498 & 270 & 203 \\
	Abs Forward Prop & 6850 & 37 & 39\\
	Imp Forward Prop & 2175 & 41 & 18\\
	\hline
	Total & 22007 & 448 & 514\\
	\hline
\end{tabular}
\end{table}

We now discuss the challenges in developing \tool.

The challenge in the automata library is 
 the usage of sets as transition labels.
Classical NFAs require only the equality comparison for transition labels.
But for s-NFAs, we have more operations for labels.
For instance, we require the operations \textbf{semI}, \textbf{intersectionI}, \textbf{nemptyI}, \textbf{memI} for  intervals.
These operations make it harder to prove the correctness of the automata library.

The challenge in the forward-propagation module is the proofs of
 Theorem \ref{thm-forward_analysis_correct}, \ref{thm-forward_sound}, and \ref{thm-forward_sound_tree}. These theorems need around 4000 lines of code to prove.

\section{Related Works}

\emph{String solvers.}
As introduced in Section \ref{sec-introduction}, there already exist various non-certified string solvers, such as Kaluza
\cite{DBLP:conf/sp/SaxenaAHMMS10},
CVC4 \cite{cvc4}, Z3 \cite{Z3}, Z3-str3 \cite{Z3-str3,z3str3re}, Z3-Trau 
\cite{Z3-trau},
S3P \cite{TCJ16}, OSTRICH \cite{CHL+19,OSTRICHlen,EMU},
SLOTH \cite{HJLRV18}, and Norn 
\cite{Abdulla14}, among many others.
These solvers are intricate and support more string operations than \tool.
Some of the solvers, such as Z3-str3 and CVC4,
opted to support more string
operations and settle with incomplete solvers (e.g. with no guarantee of termination)
that could still solve many constraints that arise in practice.
Other solvers are designed with stronger theoretical guarantees; 
for instance, OSTRICH is complete
for the straight-line string constraint fragment~\cite{DBLP:conf/popl/LinB16}.
\tool~can solve the string constraints that are out of the scope of the straight-line fragment
and guarantees to terminate for the constraints without cyclic concatenation dependencies, but it will return \emph{unknown} for constraints
that do not satisfy the tree property.

\emph{Symbolic Automata}.
\tool~depends on automata operations for regular expression propagation 
and consistency checking.
An efficient implementation of automata is crucial.
In comparison to finite-state automata in the classical sense,
symbolic automata
\cite{DBLP:conf/cav/DAntoniV17,DV21}
have proven to be more appropriate for applications like
model checking, natural language processing, and networking.
\emph{Certified automata libraries} \cite{Peter14,Julian} can provide trustworthiness
for automata-based analysis in applications.
However, to the best of our knowledge, existing certified automata 
libraries are based on the classical definition of automata,
which makes them inefficient for practical applications with
very large alphabets. Our work in this paper contributes to the development of
certified symbolic automata libraries.

\emph{Paradigms toward certified constraint solvers.}
Two main paradigms have emerged for the verification of constraint solvers:
(i)~the verification of the solver itself, i.e., the development of
solvers with machine-checked end-to-end correctness guarantees.
For instance, Shi et al. \cite{DBLP:conf/cav/ShiFLTWY20}
built a certified SMT quantifier-free bit-vector solver.
(ii)~The generation of certificates, i.e., the actual solver is not
verified, but its outputs are accompanied by proof witnesses that can then be independently checked by a
verified, trustworthy certificate checker.
For instance, Ekici et al. \cite{DBLP:journals/corr/EkiciKKMRT16}
provide an independent and certified
checker for SAT and SMT proof witnesses.
In our work, we follow the first paradigm.

\emph{Security applications based on string analysis.}
There are various security applications of automata-based analysis and
string solvers, such as detecting web security vulnerabilities.
Yu et al. \cite{DBLP:journals/fmsd/YuABI14}
investigated the approach to use automata theory for
detecting  security vulnerabilities, such as XSS, in PHP programs.
Their work also depends on a forward analysis.
But the forward analysis is over the control flow graphs of PHP programs.
\tool~aims to build a stand-alone string solver and 
the forward-propagation is only over the string constraints.
Many developments of string solvers, string analysis techniques, and
automata-theoretic techniques were also directly motivated by security 
vulnerability detection, e.g.,
\cite{DBLP:conf/sp/SaxenaAHMMS10,DBLP:conf/ccs/TrinhCJ14,DBLP:conf/popl/LinB16,popl18b,CHL+19,HJLRV18,TCJ16,aratha,DBLP:conf/fmcad/BackesBCDGLRTV18,SVB21}.
To the best of our knowledge, \tool~is the first tool that provides 
certification of a string analysis method that is applicable to 
security vulnerability detection.
We believe there is a need for further development of certified string 
analyzers: string solving techniques  are intricate and
hence error-prone, but are applied for  security analysis whose correctness is
of critical importance.

\section{Conclusion and Future Works}

In this paper, we present \tool, a certified string solver for the theory of
concatenation and regular constraints.
The backend of \tool~is verified in Isabelle proof assistant, which provides
a  rigorous guarantee for the results of the solver.
We ran \tool~over the benchmark Kaluza to show 
the efficacy of \tool.

As  future works, 
firstly,
we plan to support more string operations, such as \emph{string replacement} and
\emph{capture groups}, which are also widely used in programming languages, such
as JavaScript and PHP, increasing the applicability of \tool.
Secondly, the front end of \tool~still needs to be verified in Isabelle, especially
the correctness of the desugaring from string constraints with monadic length functions and disjunctions to the language of \tool.

\begin{acks}
	The authors would like
    to thank anonymous reviewers for their valuable comments.
	This research was supported in part by the ERC Starting Grant
	759969 (AV-SMP), Max-Planck Fellowship, Amazon Research Award,
	the Swedish Research Council (VR)
	under grant~2018-04727, and by the Swedish Foundation for Strategic
	Research (SSF) under the project WebSec (Ref.\ RIT17-0011).
\end{acks}

\balance
\bibliographystyle{ACM-Reference-Format}
\bibliography{literature}


\begin{thebibliography}{43}


\ifx \showCODEN    \undefined \def \showCODEN     #1{\unskip}     \fi
\ifx \showDOI      \undefined \def \showDOI       #1{#1}\fi
\ifx \showISBNx    \undefined \def \showISBNx     #1{\unskip}     \fi
\ifx \showISBNxiii \undefined \def \showISBNxiii  #1{\unskip}     \fi
\ifx \showISSN     \undefined \def \showISSN      #1{\unskip}     \fi
\ifx \showLCCN     \undefined \def \showLCCN      #1{\unskip}     \fi
\ifx \shownote     \undefined \def \shownote      #1{#1}          \fi
\ifx \showarticletitle \undefined \def \showarticletitle #1{#1}   \fi
\ifx \showURL      \undefined \def \showURL       {\relax}        \fi
\providecommand\bibfield[2]{#2}
\providecommand\bibinfo[2]{#2}
\providecommand\natexlab[1]{#1}
\providecommand\showeprint[2][]{arXiv:#2}

\bibitem[\protect\citeauthoryear{Abdulla, Atig, Chen, Hol{\'{\i}}k, Rezine,
  R{\"{u}}mmer, and Stenman}{Abdulla et~al\mbox{.}}{2014}]%
        {Abdulla14}
\bibfield{author}{\bibinfo{person}{Parosh~Aziz Abdulla},
  \bibinfo{person}{Mohamed~Faouzi Atig}, \bibinfo{person}{Yu{-}Fang Chen},
  \bibinfo{person}{Luk{\'{a}}s Hol{\'{\i}}k}, \bibinfo{person}{Ahmed Rezine},
  \bibinfo{person}{Philipp R{\"{u}}mmer}, {and} \bibinfo{person}{Jari
  Stenman}.} \bibinfo{year}{2014}\natexlab{}.
\newblock \showarticletitle{String Constraints for Verification}. In
  \bibinfo{booktitle}{\emph{Computer Aided Verification - 26th International
  Conference, {CAV} 2014, Held as Part of the Vienna Summer of Logic, {VSL}
  2014, Vienna, Austria, July 18-22, 2014. Proceedings}}
  \emph{(\bibinfo{series}{Lecture Notes in Computer Science},
  Vol.~\bibinfo{volume}{8559})}, \bibfield{editor}{\bibinfo{person}{Armin
  Biere} {and} \bibinfo{person}{Roderick Bloem}} (Eds.).
  \bibinfo{publisher}{Springer}, \bibinfo{pages}{150--166}.
\newblock
\urldef\tempurl%
\url{https://doi.org/10.1007/978-3-319-08867-9\_10}
\showDOI{\tempurl}


\bibitem[\protect\citeauthoryear{Amadini, Andrlon, Gange, Schachte,
  S{\o}ndergaard, and Stuckey}{Amadini et~al\mbox{.}}{2019}]%
        {aratha}
\bibfield{author}{\bibinfo{person}{Roberto Amadini}, \bibinfo{person}{Mak
  Andrlon}, \bibinfo{person}{Graeme Gange}, \bibinfo{person}{Peter Schachte},
  \bibinfo{person}{Harald S{\o}ndergaard}, {and} \bibinfo{person}{Peter~J.
  Stuckey}.} \bibinfo{year}{2019}\natexlab{}.
\newblock \showarticletitle{Constraint Programming for Dynamic Symbolic
  Execution of JavaScript}. In \bibinfo{booktitle}{\emph{Integration of
  Constraint Programming, Artificial Intelligence, and Operations Research -
  16th International Conference, {CPAIOR} 2019, Thessaloniki, Greece, June 4-7,
  2019, Proceedings}} \emph{(\bibinfo{series}{Lecture Notes in Computer
  Science}, Vol.~\bibinfo{volume}{11494})},
  \bibfield{editor}{\bibinfo{person}{Louis{-}Martin Rousseau} {and}
  \bibinfo{person}{Kostas Stergiou}} (Eds.). \bibinfo{publisher}{Springer},
  \bibinfo{pages}{1--19}.
\newblock
\urldef\tempurl%
\url{https://doi.org/10.1007/978-3-030-19212-9\_1}
\showDOI{\tempurl}


\bibitem[\protect\citeauthoryear{Backes, Bolignano, Cook, Dodge, Gacek, Luckow,
  Rungta, Tkachuk, and Varming}{Backes et~al\mbox{.}}{2018}]%
        {DBLP:conf/fmcad/BackesBCDGLRTV18}
\bibfield{author}{\bibinfo{person}{John Backes}, \bibinfo{person}{Pauline
  Bolignano}, \bibinfo{person}{Byron Cook}, \bibinfo{person}{Catherine Dodge},
  \bibinfo{person}{Andrew Gacek}, \bibinfo{person}{Kasper~S{\o}e Luckow},
  \bibinfo{person}{Neha Rungta}, \bibinfo{person}{Oksana Tkachuk}, {and}
  \bibinfo{person}{Carsten Varming}.} \bibinfo{year}{2018}\natexlab{}.
\newblock \showarticletitle{Semantic-based Automated Reasoning for {AWS} Access
  Policies using {SMT}}. In \bibinfo{booktitle}{\emph{2018 Formal Methods in
  Computer Aided Design, {FMCAD} 2018, Austin, TX, USA, October 30 - November
  2, 2018}}, \bibfield{editor}{\bibinfo{person}{Nikolaj Bj{\o}rner} {and}
  \bibinfo{person}{Arie Gurfinkel}} (Eds.). \bibinfo{publisher}{{IEEE}},
  \bibinfo{pages}{1--9}.
\newblock
\urldef\tempurl%
\url{https://doi.org/10.23919/FMCAD.2018.8602994}
\showDOI{\tempurl}


\bibitem[\protect\citeauthoryear{Berzish, Ganesh, and Zheng}{Berzish
  et~al\mbox{.}}{2017}]%
        {Z3-str3}
\bibfield{author}{\bibinfo{person}{Murphy Berzish}, \bibinfo{person}{Vijay
  Ganesh}, {and} \bibinfo{person}{Yunhui Zheng}.}
  \bibinfo{year}{2017}\natexlab{}.
\newblock \showarticletitle{Z3str3: {A} string solver with theory-aware
  heuristics}. In \bibinfo{booktitle}{\emph{2017 Formal Methods in Computer
  Aided Design, {FMCAD} 2017, Vienna, Austria, October 2-6, 2017}},
  \bibfield{editor}{\bibinfo{person}{Daryl Stewart} {and}
  \bibinfo{person}{Georg Weissenbacher}} (Eds.). \bibinfo{publisher}{{IEEE}},
  \bibinfo{pages}{55--59}.
\newblock
\urldef\tempurl%
\url{https://doi.org/10.23919/FMCAD.2017.8102241}
\showDOI{\tempurl}


\bibitem[\protect\citeauthoryear{Berzish, Kulczynski, Mora, Manea, Day,
  Nowotka, and Ganesh}{Berzish et~al\mbox{.}}{2021}]%
        {z3str3re}
\bibfield{author}{\bibinfo{person}{Murphy Berzish}, \bibinfo{person}{Mitja
  Kulczynski}, \bibinfo{person}{Federico Mora}, \bibinfo{person}{Florin Manea},
  \bibinfo{person}{Joel~D. Day}, \bibinfo{person}{Dirk Nowotka}, {and}
  \bibinfo{person}{Vijay Ganesh}.} \bibinfo{year}{2021}\natexlab{}.
\newblock \showarticletitle{An {SMT} Solver for Regular Expressions and Linear
  Arithmetic over String Length}. In \bibinfo{booktitle}{\emph{Computer Aided
  Verification - 33rd International Conference, {CAV} 2021, Virtual Event, July
  20-23, 2021, Proceedings, Part {II}}} \emph{(\bibinfo{series}{Lecture Notes
  in Computer Science}, Vol.~\bibinfo{volume}{12760})},
  \bibfield{editor}{\bibinfo{person}{Alexandra Silva} {and}
  \bibinfo{person}{K.~Rustan~M. Leino}} (Eds.). \bibinfo{publisher}{Springer},
  \bibinfo{pages}{289--312}.
\newblock
\urldef\tempurl%
\url{https://doi.org/10.1007/978-3-030-81688-9\_14}
\showDOI{\tempurl}


\bibitem[\protect\citeauthoryear{Blotsky, Mora, Berzish, Zheng, Kabir, and
  Ganesh}{Blotsky et~al\mbox{.}}{2018}]%
        {DBLP:conf/cav/BlotskyMBZKG18}
\bibfield{author}{\bibinfo{person}{Dmitry Blotsky}, \bibinfo{person}{Federico
  Mora}, \bibinfo{person}{Murphy Berzish}, \bibinfo{person}{Yunhui Zheng},
  \bibinfo{person}{Ifaz Kabir}, {and} \bibinfo{person}{Vijay Ganesh}.}
  \bibinfo{year}{2018}\natexlab{}.
\newblock \showarticletitle{StringFuzz: {A} Fuzzer for String Solvers}. In
  \bibinfo{booktitle}{\emph{Computer Aided Verification - 30th International
  Conference, {CAV} 2018, Held as Part of the Federated Logic Conference, FloC
  2018, Oxford, UK, July 14-17, 2018, Proceedings, Part {II}}}
  \emph{(\bibinfo{series}{Lecture Notes in Computer Science},
  Vol.~\bibinfo{volume}{10982})}, \bibfield{editor}{\bibinfo{person}{Hana
  Chockler} {and} \bibinfo{person}{Georg Weissenbacher}} (Eds.).
  \bibinfo{publisher}{Springer}, \bibinfo{pages}{45--51}.
\newblock
\urldef\tempurl%
\url{https://doi.org/10.1007/978-3-319-96142-2\_6}
\showDOI{\tempurl}


\bibitem[\protect\citeauthoryear{Brunner}{Brunner}{2017}]%
        {Peter14}
\bibfield{author}{\bibinfo{person}{Julian Brunner}.}
  \bibinfo{year}{2017}\natexlab{}.
\newblock \bibinfo{title}{{Transition Systems and Automata Isabelle Library}}.
\newblock \bibinfo{howpublished}{Arch. Formal Proofs}.
\newblock
\newblock
\shownote{\url{https://www.isa-afp.org/entries/Transition_Systems_and_Automata.html}.}


\bibitem[\protect\citeauthoryear{Bugariu and M{\"{u}}ller}{Bugariu and
  M{\"{u}}ller}{2020}]%
        {BM20}
\bibfield{author}{\bibinfo{person}{Alexandra Bugariu} {and}
  \bibinfo{person}{Peter M{\"{u}}ller}.} \bibinfo{year}{2020}\natexlab{}.
\newblock \showarticletitle{Automatically testing string solvers}. In
  \bibinfo{booktitle}{\emph{{ICSE} '20: 42nd International Conference on
  Software Engineering, Seoul, South Korea, 27 June - 19 July, 2020}},
  \bibfield{editor}{\bibinfo{person}{Gregg Rothermel} {and}
  \bibinfo{person}{Doo{-}Hwan Bae}} (Eds.). \bibinfo{publisher}{{ACM}},
  \bibinfo{pages}{1459--1470}.
\newblock
\urldef\tempurl%
\url{https://doi.org/10.1145/3377811.3380398}
\showDOI{\tempurl}


\bibitem[\protect\citeauthoryear{Bui and contributors}{Bui and
  contributors}{2019}]%
        {Z3-trau}
\bibfield{author}{\bibinfo{person}{Diep Bui} {and}
  \bibinfo{person}{contributors}.} \bibinfo{year}{2019}\natexlab{}.
\newblock \bibinfo{title}{Z3-Trau}.
\newblock \bibinfo{howpublished}{\url{https://github.com/diepbp/z3-trau}}.
\newblock


\bibitem[\protect\citeauthoryear{Chen, Chen, Hague, Lin, and Wu}{Chen
  et~al\mbox{.}}{2018}]%
        {popl18b}
\bibfield{author}{\bibinfo{person}{Taolue Chen}, \bibinfo{person}{Yan Chen},
  \bibinfo{person}{Matthew Hague}, \bibinfo{person}{Anthony~W. Lin}, {and}
  \bibinfo{person}{Zhilin Wu}.} \bibinfo{year}{2018}\natexlab{}.
\newblock \showarticletitle{What is decidable about string constraints with the
  ReplaceAll function}.
\newblock \bibinfo{journal}{\emph{Proc. {ACM} Program. Lang.}}
  \bibinfo{volume}{2}, \bibinfo{number}{{POPL}} (\bibinfo{year}{2018}),
  \bibinfo{pages}{3:1--3:29}.
\newblock
\urldef\tempurl%
\url{https://doi.org/10.1145/3158091}
\showDOI{\tempurl}


\bibitem[\protect\citeauthoryear{Chen, Flores-Lamas, Hague, Han, Hu, Kan, Lin,
  Ruemmer, and Wu}{Chen et~al\mbox{.}}{2022}]%
        {EMU}
\bibfield{author}{\bibinfo{person}{Taolue Chen}, \bibinfo{person}{Alejandro
  Flores-Lamas}, \bibinfo{person}{Matthew Hague}, \bibinfo{person}{Zhilei Han},
  \bibinfo{person}{Denghang Hu}, \bibinfo{person}{Shuanglong Kan},
  \bibinfo{person}{Anthony~W. Lin}, \bibinfo{person}{Philipp Ruemmer}, {and}
  \bibinfo{person}{Zhilin Wu}.} \bibinfo{year}{2022}\natexlab{}.
\newblock \showarticletitle{Solving String Constraints with Regex-Dependent
  Functions through Transducers with Priorities and Variables}.
\newblock \bibinfo{journal}{\emph{Proc. {ACM} Program. Lang.}}
  \bibinfo{volume}{6}, \bibinfo{number}{{POPL}} (\bibinfo{year}{2022}).
\newblock


\bibitem[\protect\citeauthoryear{Chen, Hague, He, Hu, Lin, R{\"{u}}mmer, and
  Wu}{Chen et~al\mbox{.}}{2020}]%
        {OSTRICHlen}
\bibfield{author}{\bibinfo{person}{Taolue Chen}, \bibinfo{person}{Matthew
  Hague}, \bibinfo{person}{Jinlong He}, \bibinfo{person}{Denghang Hu},
  \bibinfo{person}{Anthony~Widjaja Lin}, \bibinfo{person}{Philipp
  R{\"{u}}mmer}, {and} \bibinfo{person}{Zhilin Wu}.}
  \bibinfo{year}{2020}\natexlab{}.
\newblock \showarticletitle{A Decision Procedure for Path Feasibility of String
  Manipulating Programs with Integer Data Type}. In
  \bibinfo{booktitle}{\emph{Automated Technology for Verification and Analysis
  - 18th International Symposium, {ATVA} 2020, Hanoi, Vietnam, October 19-23,
  2020, Proceedings}}. \bibinfo{pages}{325--342}.
\newblock
\urldef\tempurl%
\url{https://doi.org/10.1007/978-3-030-59152-6\_18}
\showDOI{\tempurl}


\bibitem[\protect\citeauthoryear{Chen, Hague, Lin, R{\"{u}}mmer, and Wu}{Chen
  et~al\mbox{.}}{2019}]%
        {CHL+19}
\bibfield{author}{\bibinfo{person}{Taolue Chen}, \bibinfo{person}{Matthew
  Hague}, \bibinfo{person}{Anthony~W. Lin}, \bibinfo{person}{Philipp
  R{\"{u}}mmer}, {and} \bibinfo{person}{Zhilin Wu}.}
  \bibinfo{year}{2019}\natexlab{}.
\newblock \showarticletitle{Decision procedures for path feasibility of
  string-manipulating programs with complex operations}.
\newblock \bibinfo{journal}{\emph{Proc. {ACM} Program. Lang.}}
  \bibinfo{volume}{3}, \bibinfo{number}{{POPL}} (\bibinfo{year}{2019}),
  \bibinfo{pages}{49:1--49:30}.
\newblock
\urldef\tempurl%
\url{https://doi.org/10.1145/3290362}
\showDOI{\tempurl}


\bibitem[\protect\citeauthoryear{Cordeiro, Kesseli, Kroening, Schrammel, and
  Trt{\'{\i}}k}{Cordeiro et~al\mbox{.}}{2018}]%
        {DBLP:conf/cav/CordeiroKKST18}
\bibfield{author}{\bibinfo{person}{Lucas~C. Cordeiro}, \bibinfo{person}{Pascal
  Kesseli}, \bibinfo{person}{Daniel Kroening}, \bibinfo{person}{Peter
  Schrammel}, {and} \bibinfo{person}{Marek Trt{\'{\i}}k}.}
  \bibinfo{year}{2018}\natexlab{}.
\newblock \showarticletitle{{JBMC:} {A} Bounded Model Checking Tool for
  Verifying Java Bytecode}. In \bibinfo{booktitle}{\emph{Computer Aided
  Verification - 30th International Conference, {CAV} 2018, Held as Part of the
  Federated Logic Conference, FloC 2018, Oxford, UK, July 14-17, 2018,
  Proceedings, Part {I}}} \emph{(\bibinfo{series}{Lecture Notes in Computer
  Science}, Vol.~\bibinfo{volume}{10981})},
  \bibfield{editor}{\bibinfo{person}{Hana Chockler} {and}
  \bibinfo{person}{Georg Weissenbacher}} (Eds.). \bibinfo{publisher}{Springer},
  \bibinfo{pages}{183--190}.
\newblock
\urldef\tempurl%
\url{https://doi.org/10.1007/978-3-319-96145-3\_10}
\showDOI{\tempurl}


\bibitem[\protect\citeauthoryear{D'Antoni and Veanes}{D'Antoni and
  Veanes}{2017}]%
        {DBLP:conf/cav/DAntoniV17}
\bibfield{author}{\bibinfo{person}{Loris D'Antoni} {and}
  \bibinfo{person}{Margus Veanes}.} \bibinfo{year}{2017}\natexlab{}.
\newblock \showarticletitle{The Power of Symbolic Automata and Transducers}. In
  \bibinfo{booktitle}{\emph{Computer Aided Verification - 29th International
  Conference, {CAV} 2017, Heidelberg, Germany, July 24-28, 2017, Proceedings,
  Part {I}}} \emph{(\bibinfo{series}{Lecture Notes in Computer Science},
  Vol.~\bibinfo{volume}{10426})}, \bibfield{editor}{\bibinfo{person}{Rupak
  Majumdar} {and} \bibinfo{person}{Viktor Kuncak}} (Eds.).
  \bibinfo{publisher}{Springer}, \bibinfo{pages}{47--67}.
\newblock
\urldef\tempurl%
\url{https://doi.org/10.1007/978-3-319-63387-9\_3}
\showDOI{\tempurl}


\bibitem[\protect\citeauthoryear{D'Antoni and Veanes}{D'Antoni and
  Veanes}{2021}]%
        {DV21}
\bibfield{author}{\bibinfo{person}{Loris D'Antoni} {and}
  \bibinfo{person}{Margus Veanes}.} \bibinfo{year}{2021}\natexlab{}.
\newblock \showarticletitle{Automata modulo theories}.
\newblock \bibinfo{journal}{\emph{Commun. {ACM}}} \bibinfo{volume}{64},
  \bibinfo{number}{5} (\bibinfo{year}{2021}), \bibinfo{pages}{86--95}.
\newblock
\urldef\tempurl%
\url{https://doi.org/10.1145/3419404}
\showDOI{\tempurl}


\bibitem[\protect\citeauthoryear{de~Moura and Bj{\o}rner}{de~Moura and
  Bj{\o}rner}{2008}]%
        {Z3}
\bibfield{author}{\bibinfo{person}{Leonardo~Mendon{\c{c}}a de Moura} {and}
  \bibinfo{person}{Nikolaj Bj{\o}rner}.} \bibinfo{year}{2008}\natexlab{}.
\newblock \showarticletitle{{Z3:} An Efficient {SMT} Solver}. In
  \bibinfo{booktitle}{\emph{Tools and Algorithms for the Construction and
  Analysis of Systems, 14th International Conference, {TACAS} 2008, Held as
  Part of the Joint European Conferences on Theory and Practice of Software,
  {ETAPS} 2008, Budapest, Hungary, March 29-April 6, 2008. Proceedings}}
  \emph{(\bibinfo{series}{Lecture Notes in Computer Science},
  Vol.~\bibinfo{volume}{4963})}, \bibfield{editor}{\bibinfo{person}{C.~R.
  Ramakrishnan} {and} \bibinfo{person}{Jakob Rehof}} (Eds.).
  \bibinfo{publisher}{Springer}, \bibinfo{pages}{337--340}.
\newblock
\urldef\tempurl%
\url{https://doi.org/10.1007/978-3-540-78800-3\_24}
\showDOI{\tempurl}


\bibitem[\protect\citeauthoryear{Diekert}{Diekert}{2002}]%
        {Diekert}
\bibfield{author}{\bibinfo{person}{Volker Diekert}.}
  \bibinfo{year}{2002}\natexlab{}.
\newblock \showarticletitle{Makanin's {A}lgorithm}.
\newblock In \bibinfo{booktitle}{\emph{Algebraic Combinatorics on Words}},
  \bibfield{editor}{\bibinfo{person}{M.~Lothaire}} (Ed.).
  \bibinfo{series}{Encyclopedia of Mathematics and its Applications},
  Vol.~\bibinfo{volume}{90}. \bibinfo{publisher}{Cambridge University Press},
  Chapter~12, \bibinfo{pages}{387--442}.
\newblock


\bibitem[\protect\citeauthoryear{Ekici, Katz, Keller, Mebsout, Reynolds, and
  Tinelli}{Ekici et~al\mbox{.}}{2016}]%
        {DBLP:journals/corr/EkiciKKMRT16}
\bibfield{author}{\bibinfo{person}{Burak Ekici}, \bibinfo{person}{Guy Katz},
  \bibinfo{person}{Chantal Keller}, \bibinfo{person}{Alain Mebsout},
  \bibinfo{person}{Andrew~J. Reynolds}, {and} \bibinfo{person}{Cesare
  Tinelli}.} \bibinfo{year}{2016}\natexlab{}.
\newblock \showarticletitle{Extending SMTCoq, a Certified Checker for {SMT}
  (Extended Abstract)}. In \bibinfo{booktitle}{\emph{Proceedings First
  International Workshop on Hammers for Type Theories, HaTT@IJCAR 2016,
  Coimbra, Portugal, July 1, 2016}} \emph{(\bibinfo{series}{{EPTCS}},
  Vol.~\bibinfo{volume}{210})},
  \bibfield{editor}{\bibinfo{person}{Jasmin~Christian Blanchette} {and}
  \bibinfo{person}{Cezary Kaliszyk}} (Eds.). \bibinfo{pages}{21--29}.
\newblock
\urldef\tempurl%
\url{https://doi.org/10.4204/EPTCS.210.5}
\showDOI{\tempurl}


\bibitem[\protect\citeauthoryear{Guti{\'{e}}rrez}{Guti{\'{e}}rrez}{1998}]%
        {Gut98}
\bibfield{author}{\bibinfo{person}{Claudio Guti{\'{e}}rrez}.}
  \bibinfo{year}{1998}\natexlab{}.
\newblock \showarticletitle{Solving Equations in Strings: On Makanin's
  Algorithm}. In \bibinfo{booktitle}{\emph{{LATIN} '98: Theoretical
  Informatics, Third Latin American Symposium, Campinas, Brazil, April, 20-24,
  1998, Proceedings}} \emph{(\bibinfo{series}{Lecture Notes in Computer
  Science}, Vol.~\bibinfo{volume}{1380})},
  \bibfield{editor}{\bibinfo{person}{Claudio~L. Lucchesi} {and}
  \bibinfo{person}{Arnaldo~V. Moura}} (Eds.). \bibinfo{publisher}{Springer},
  \bibinfo{pages}{358--373}.
\newblock
\urldef\tempurl%
\url{https://doi.org/10.1007/BFb0054336}
\showDOI{\tempurl}


\bibitem[\protect\citeauthoryear{Hague, Lin, R{\"{u}}mmer, and Wu}{Hague
  et~al\mbox{.}}{2020}]%
        {HLRW20}
\bibfield{author}{\bibinfo{person}{Matthew Hague}, \bibinfo{person}{Anthony~W.
  Lin}, \bibinfo{person}{Philipp R{\"{u}}mmer}, {and} \bibinfo{person}{Zhilin
  Wu}.} \bibinfo{year}{2020}\natexlab{}.
\newblock \showarticletitle{Monadic Decomposition in Integer Linear
  Arithmetic}. In \bibinfo{booktitle}{\emph{Automated Reasoning - 10th
  International Joint Conference, {IJCAR} 2020, Paris, France, July 1-4, 2020,
  Proceedings, Part {I}}} \emph{(\bibinfo{series}{Lecture Notes in Computer
  Science}, Vol.~\bibinfo{volume}{12166})},
  \bibfield{editor}{\bibinfo{person}{Nicolas Peltier} {and}
  \bibinfo{person}{Viorica Sofronie{-}Stokkermans}} (Eds.).
  \bibinfo{publisher}{Springer}, \bibinfo{pages}{122--140}.
\newblock
\urldef\tempurl%
\url{https://doi.org/10.1007/978-3-030-51074-9\_8}
\showDOI{\tempurl}


\bibitem[\protect\citeauthoryear{Hol{\'{\i}}k, Janku, Lin, R{\"{u}}mmer, and
  Vojnar}{Hol{\'{\i}}k et~al\mbox{.}}{2018}]%
        {HJLRV18}
\bibfield{author}{\bibinfo{person}{Luk{\'{a}}s Hol{\'{\i}}k},
  \bibinfo{person}{Petr Janku}, \bibinfo{person}{Anthony~W. Lin},
  \bibinfo{person}{Philipp R{\"{u}}mmer}, {and} \bibinfo{person}{Tom{\'{a}}s
  Vojnar}.} \bibinfo{year}{2018}\natexlab{}.
\newblock \showarticletitle{String constraints with concatenation and
  transducers solved efficiently}.
\newblock \bibinfo{journal}{\emph{Proc. {ACM} Program. Lang.}}
  \bibinfo{volume}{2}, \bibinfo{number}{{POPL}} (\bibinfo{year}{2018}),
  \bibinfo{pages}{4:1--4:32}.
\newblock
\urldef\tempurl%
\url{https://doi.org/10.1145/3158092}
\showDOI{\tempurl}


\bibitem[\protect\citeauthoryear{Jez}{Jez}{2016}]%
        {J16}
\bibfield{author}{\bibinfo{person}{Artur Jez}.}
  \bibinfo{year}{2016}\natexlab{}.
\newblock \showarticletitle{Recompression: {A} Simple and Powerful Technique
  for Word Equations}.
\newblock \bibinfo{journal}{\emph{J. {ACM}}} \bibinfo{volume}{63},
  \bibinfo{number}{1} (\bibinfo{year}{2016}), \bibinfo{pages}{4:1--4:51}.
\newblock
\urldef\tempurl%
\url{https://doi.org/10.1145/2743014}
\showDOI{\tempurl}


\bibitem[\protect\citeauthoryear{Klarlund, M{\o}ller, and
  Schwartzbach}{Klarlund et~al\mbox{.}}{2002}]%
        {mona-secret}
\bibfield{author}{\bibinfo{person}{Nils Klarlund}, \bibinfo{person}{Anders
  M{\o}ller}, {and} \bibinfo{person}{Michael~I. Schwartzbach}.}
  \bibinfo{year}{2002}\natexlab{}.
\newblock \showarticletitle{{MONA} Implementation Secrets}.
\newblock \bibinfo{journal}{\emph{Int. J. Found. Comput. Sci.}}
  \bibinfo{volume}{13}, \bibinfo{number}{4} (\bibinfo{year}{2002}),
  \bibinfo{pages}{571--586}.
\newblock
\urldef\tempurl%
\url{https://doi.org/10.1142/S012905410200128X}
\showDOI{\tempurl}


\bibitem[\protect\citeauthoryear{Lammich}{Lammich}{2013}]%
        {DBLP:conf/itp/Lammich13}
\bibfield{author}{\bibinfo{person}{Peter Lammich}.}
  \bibinfo{year}{2013}\natexlab{}.
\newblock \showarticletitle{{Automatic Data Refinement}}. In
  \bibinfo{booktitle}{\emph{Interactive Theorem Proving - 4th International
  Conference, {ITP} 2013, Rennes, France, July 22-26, 2013. Proceedings}}
  \emph{(\bibinfo{series}{Lecture Notes in Computer Science},
  Vol.~\bibinfo{volume}{7998})}, \bibfield{editor}{\bibinfo{person}{Sandrine
  Blazy}, \bibinfo{person}{Christine Paulin{-}Mohring}, {and}
  \bibinfo{person}{David Pichardie}} (Eds.). \bibinfo{publisher}{Springer},
  \bibinfo{pages}{84--99}.
\newblock
\urldef\tempurl%
\url{https://doi.org/10.1007/978-3-642-39634-2\_9}
\showDOI{\tempurl}


\bibitem[\protect\citeauthoryear{Lammich}{Lammich}{2014}]%
        {Julian}
\bibfield{author}{\bibinfo{person}{Peter Lammich}.}
  \bibinfo{year}{2014}\natexlab{}.
\newblock \bibinfo{title}{{The CAVA Automata Isabelle Library}}.
\newblock \bibinfo{howpublished}{Arch. Formal Proofs}.
\newblock
\newblock
\shownote{\url{https://www.isa-afp.org/entries/CAVA_Automata.html}.}


\bibitem[\protect\citeauthoryear{Lammich and Lochbihler}{Lammich and
  Lochbihler}{2010}]%
        {DBLP:conf/itp/LammichL10}
\bibfield{author}{\bibinfo{person}{Peter Lammich} {and}
  \bibinfo{person}{Andreas Lochbihler}.} \bibinfo{year}{2010}\natexlab{}.
\newblock \showarticletitle{{The Isabelle Collections Framework}}. In
  \bibinfo{booktitle}{\emph{Interactive Theorem Proving, First International
  Conference, {ITP} 2010, Edinburgh, UK, July 11-14, 2010. Proceedings}}
  \emph{(\bibinfo{series}{Lecture Notes in Computer Science},
  Vol.~\bibinfo{volume}{6172})}, \bibfield{editor}{\bibinfo{person}{Matt
  Kaufmann} {and} \bibinfo{person}{Lawrence~C. Paulson}} (Eds.).
  \bibinfo{publisher}{Springer}, \bibinfo{pages}{339--354}.
\newblock
\urldef\tempurl%
\url{https://doi.org/10.1007/978-3-642-14052-5\_24}
\showDOI{\tempurl}


\bibitem[\protect\citeauthoryear{Liang, Reynolds, Tinelli, Barrett, and
  Deters}{Liang et~al\mbox{.}}{2014}]%
        {cvc4}
\bibfield{author}{\bibinfo{person}{Tianyi Liang}, \bibinfo{person}{Andrew
  Reynolds}, \bibinfo{person}{Cesare Tinelli}, \bibinfo{person}{Clark~W.
  Barrett}, {and} \bibinfo{person}{Morgan Deters}.}
  \bibinfo{year}{2014}\natexlab{}.
\newblock \showarticletitle{A {DPLL(T)} Theory Solver for a Theory of Strings
  and Regular Expressions}. In \bibinfo{booktitle}{\emph{Computer Aided
  Verification - 26th International Conference, {CAV} 2014, Held as Part of the
  Vienna Summer of Logic, {VSL} 2014, Vienna, Austria, July 18-22, 2014.
  Proceedings}} \emph{(\bibinfo{series}{Lecture Notes in Computer Science},
  Vol.~\bibinfo{volume}{8559})}, \bibfield{editor}{\bibinfo{person}{Armin
  Biere} {and} \bibinfo{person}{Roderick Bloem}} (Eds.).
  \bibinfo{publisher}{Springer}, \bibinfo{pages}{646--662}.
\newblock
\urldef\tempurl%
\url{https://doi.org/10.1007/978-3-319-08867-9\_43}
\showDOI{\tempurl}


\bibitem[\protect\citeauthoryear{Lin and Barcel{\'{o}}}{Lin and
  Barcel{\'{o}}}{2016}]%
        {DBLP:conf/popl/LinB16}
\bibfield{author}{\bibinfo{person}{Anthony~Widjaja Lin} {and}
  \bibinfo{person}{Pablo Barcel{\'{o}}}.} \bibinfo{year}{2016}\natexlab{}.
\newblock \showarticletitle{String solving with word equations and transducers:
  towards a logic for analysing mutation {XSS}}. In
  \bibinfo{booktitle}{\emph{Proceedings of the 43rd Annual {ACM}
  {SIGPLAN-SIGACT} Symposium on Principles of Programming Languages, {POPL}
  2016, St. Petersburg, FL, USA, January 20 - 22, 2016}},
  \bibfield{editor}{\bibinfo{person}{Rastislav Bod{\'{\i}}k} {and}
  \bibinfo{person}{Rupak Majumdar}} (Eds.). \bibinfo{publisher}{{ACM}},
  \bibinfo{pages}{123--136}.
\newblock
\urldef\tempurl%
\url{https://doi.org/10.1145/2837614.2837641}
\showDOI{\tempurl}


\bibitem[\protect\citeauthoryear{Makanin}{Makanin}{1977}]%
        {Makanin}
\bibfield{author}{\bibinfo{person}{Gennady~S Makanin}.}
  \bibinfo{year}{1977}\natexlab{}.
\newblock \showarticletitle{The problem of solvability of equations in a free
  semigroup}.
\newblock \bibinfo{journal}{\emph{Sbornik: Mathematics}} \bibinfo{volume}{32},
  \bibinfo{number}{2} (\bibinfo{year}{1977}), \bibinfo{pages}{129--198}.
\newblock


\bibitem[\protect\citeauthoryear{Mansur, Christakis, W{\"{u}}stholz, and
  Zhang}{Mansur et~al\mbox{.}}{2020}]%
        {Mansur20}
\bibfield{author}{\bibinfo{person}{Muhammad~Numair Mansur},
  \bibinfo{person}{Maria Christakis}, \bibinfo{person}{Valentin
  W{\"{u}}stholz}, {and} \bibinfo{person}{Fuyuan Zhang}.}
  \bibinfo{year}{2020}\natexlab{}.
\newblock \showarticletitle{Detecting critical bugs in {SMT} solvers using
  blackbox mutational fuzzing}. In \bibinfo{booktitle}{\emph{{ESEC/FSE} '20:
  28th {ACM} Joint European Software Engineering Conference and Symposium on
  the Foundations of Software Engineering, Virtual Event, USA, November 8-13,
  2020}}, \bibfield{editor}{\bibinfo{person}{Prem Devanbu},
  \bibinfo{person}{Myra~B. Cohen}, {and} \bibinfo{person}{Thomas Zimmermann}}
  (Eds.). \bibinfo{publisher}{{ACM}}, \bibinfo{pages}{701--712}.
\newblock
\urldef\tempurl%
\url{https://doi.org/10.1145/3368089.3409763}
\showDOI{\tempurl}


\bibitem[\protect\citeauthoryear{Minamide}{Minamide}{2005}]%
        {Min05}
\bibfield{author}{\bibinfo{person}{Yasuhiko Minamide}.}
  \bibinfo{year}{2005}\natexlab{}.
\newblock \showarticletitle{Static approximation of dynamically generated Web
  pages}. In \bibinfo{booktitle}{\emph{Proceedings of the 14th international
  conference on World Wide Web, {WWW} 2005, Chiba, Japan, May 10-14, 2005}},
  \bibfield{editor}{\bibinfo{person}{Allan Ellis} {and}
  \bibinfo{person}{Tatsuya Hagino}} (Eds.). \bibinfo{publisher}{{ACM}},
  \bibinfo{pages}{432--441}.
\newblock
\urldef\tempurl%
\url{https://doi.org/10.1145/1060745.1060809}
\showDOI{\tempurl}


\bibitem[\protect\citeauthoryear{Nipkow, Paulson, and Wenzel}{Nipkow
  et~al\mbox{.}}{2002}]%
        {nipkow2002isabelle}
\bibfield{author}{\bibinfo{person}{Tobias Nipkow}, \bibinfo{person}{Lawrence~C
  Paulson}, {and} \bibinfo{person}{Markus Wenzel}.}
  \bibinfo{year}{2002}\natexlab{}.
\newblock \bibinfo{booktitle}{\emph{Isabelle/HOL: a proof assistant for
  higher-order logic}}. Vol.~\bibinfo{volume}{2283}.
\newblock \bibinfo{publisher}{Springer Science \& Business Media}.
\newblock


\bibitem[\protect\citeauthoryear{Noller, Pasareanu, Fromherz, Le, and
  Visser}{Noller et~al\mbox{.}}{2019}]%
        {DBLP:conf/tacas/NollerPFLV19}
\bibfield{author}{\bibinfo{person}{Yannic Noller}, \bibinfo{person}{Corina~S.
  Pasareanu}, \bibinfo{person}{Aymeric Fromherz},
  \bibinfo{person}{Xuan{-}Bach~Dinh Le}, {and} \bibinfo{person}{Willem
  Visser}.} \bibinfo{year}{2019}\natexlab{}.
\newblock \showarticletitle{Symbolic Pathfinder for {SV-COMP} - (Competition
  Contribution)}. In \bibinfo{booktitle}{\emph{Tools and Algorithms for the
  Construction and Analysis of Systems - 25 Years of {TACAS:} TOOLympics, Held
  as Part of {ETAPS} 2019, Prague, Czech Republic, April 6-11, 2019,
  Proceedings, Part {III}}} \emph{(\bibinfo{series}{Lecture Notes in Computer
  Science}, Vol.~\bibinfo{volume}{11429})},
  \bibfield{editor}{\bibinfo{person}{Dirk Beyer}, \bibinfo{person}{Marieke
  Huisman}, \bibinfo{person}{Fabrice Kordon}, {and} \bibinfo{person}{Bernhard
  Steffen}} (Eds.). \bibinfo{publisher}{Springer}, \bibinfo{pages}{239--243}.
\newblock
\urldef\tempurl%
\url{https://doi.org/10.1007/978-3-030-17502-3\_21}
\showDOI{\tempurl}


\bibitem[\protect\citeauthoryear{Saxena, Akhawe, Hanna, Mao, McCamant, and
  Song}{Saxena et~al\mbox{.}}{2010}]%
        {DBLP:conf/sp/SaxenaAHMMS10}
\bibfield{author}{\bibinfo{person}{Prateek Saxena}, \bibinfo{person}{Devdatta
  Akhawe}, \bibinfo{person}{Steve Hanna}, \bibinfo{person}{Feng Mao},
  \bibinfo{person}{Stephen McCamant}, {and} \bibinfo{person}{Dawn Song}.}
  \bibinfo{year}{2010}\natexlab{}.
\newblock \showarticletitle{A Symbolic Execution Framework for JavaScript}. In
  \bibinfo{booktitle}{\emph{31st {IEEE} Symposium on Security and Privacy,
  S{\&}P 2010, 16-19 May 2010, Berleley/Oakland, California, {USA}}}.
  \bibinfo{publisher}{{IEEE} Computer Society}, \bibinfo{pages}{513--528}.
\newblock
\urldef\tempurl%
\url{https://doi.org/10.1109/SP.2010.38}
\showDOI{\tempurl}


\bibitem[\protect\citeauthoryear{Shi, Fu, Liu, Tsai, Wang, and Yang}{Shi
  et~al\mbox{.}}{2021}]%
        {DBLP:conf/cav/ShiFLTWY20}
\bibfield{author}{\bibinfo{person}{Xiaomu Shi}, \bibinfo{person}{Yu{-}Fu Fu},
  \bibinfo{person}{Jiaxiang Liu}, \bibinfo{person}{Ming{-}Hsien Tsai},
  \bibinfo{person}{Bow{-}Yaw Wang}, {and} \bibinfo{person}{Bo{-}Yin Yang}.}
  \bibinfo{year}{2021}\natexlab{}.
\newblock \showarticletitle{CoqQFBV: {A} Scalable Certified {SMT}
  Quantifier-Free Bit-Vector Solver}. In \bibinfo{booktitle}{\emph{Computer
  Aided Verification - 33rd International Conference, {CAV} 2021, Virtual
  Event, July 20-23, 2021, Proceedings, Part {II}}}
  \emph{(\bibinfo{series}{Lecture Notes in Computer Science},
  Vol.~\bibinfo{volume}{12760})}, \bibfield{editor}{\bibinfo{person}{Alexandra
  Silva} {and} \bibinfo{person}{K.~Rustan~M. Leino}} (Eds.).
  \bibinfo{publisher}{Springer}, \bibinfo{pages}{149--171}.
\newblock
\urldef\tempurl%
\url{https://doi.org/10.1007/978-3-030-81688-9\_7}
\showDOI{\tempurl}


\bibitem[\protect\citeauthoryear{Stanford, Veanes, and Bj{\o}rner}{Stanford
  et~al\mbox{.}}{2021}]%
        {SVB21}
\bibfield{author}{\bibinfo{person}{Caleb Stanford}, \bibinfo{person}{Margus
  Veanes}, {and} \bibinfo{person}{Nikolaj Bj{\o}rner}.}
  \bibinfo{year}{2021}\natexlab{}.
\newblock \showarticletitle{Symbolic Boolean derivatives for efficiently
  solving extended regular expression constraints}. In
  \bibinfo{booktitle}{\emph{{PLDI} '21: 42nd {ACM} {SIGPLAN} International
  Conference on Programming Language Design and Implementation, Virtual Event,
  Canada, June 20-25, 2021}}, \bibfield{editor}{\bibinfo{person}{Stephen~N.
  Freund} {and} \bibinfo{person}{Eran Yahav}} (Eds.).
  \bibinfo{publisher}{{ACM}}, \bibinfo{pages}{620--635}.
\newblock
\urldef\tempurl%
\url{https://doi.org/10.1145/3453483.3454066}
\showDOI{\tempurl}


\bibitem[\protect\citeauthoryear{Trinh, Chu, and Jaffar}{Trinh
  et~al\mbox{.}}{2014}]%
        {DBLP:conf/ccs/TrinhCJ14}
\bibfield{author}{\bibinfo{person}{Minh{-}Thai Trinh},
  \bibinfo{person}{Duc{-}Hiep Chu}, {and} \bibinfo{person}{Joxan Jaffar}.}
  \bibinfo{year}{2014}\natexlab{}.
\newblock \showarticletitle{{S3:} {A} Symbolic String Solver for Vulnerability
  Detection in Web Applications}. In \bibinfo{booktitle}{\emph{Proceedings of
  the 2014 {ACM} {SIGSAC} Conference on Computer and Communications Security,
  Scottsdale, AZ, USA, November 3-7, 2014}},
  \bibfield{editor}{\bibinfo{person}{Gail{-}Joon Ahn}, \bibinfo{person}{Moti
  Yung}, {and} \bibinfo{person}{Ninghui Li}} (Eds.).
  \bibinfo{publisher}{{ACM}}, \bibinfo{pages}{1232--1243}.
\newblock
\urldef\tempurl%
\url{https://doi.org/10.1145/2660267.2660372}
\showDOI{\tempurl}


\bibitem[\protect\citeauthoryear{Trinh, Chu, and Jaffar}{Trinh
  et~al\mbox{.}}{2016}]%
        {TCJ16}
\bibfield{author}{\bibinfo{person}{Minh{-}Thai Trinh},
  \bibinfo{person}{Duc{-}Hiep Chu}, {and} \bibinfo{person}{Joxan Jaffar}.}
  \bibinfo{year}{2016}\natexlab{}.
\newblock \showarticletitle{Progressive Reasoning over Recursively-Defined
  Strings}. In \bibinfo{booktitle}{\emph{Computer Aided Verification - 28th
  International Conference, {CAV} 2016, Toronto, ON, Canada, July 17-23, 2016,
  Proceedings, Part {I}}} \emph{(\bibinfo{series}{Lecture Notes in Computer
  Science}, Vol.~\bibinfo{volume}{9779})},
  \bibfield{editor}{\bibinfo{person}{Swarat Chaudhuri} {and}
  \bibinfo{person}{Azadeh Farzan}} (Eds.). \bibinfo{publisher}{Springer},
  \bibinfo{pages}{218--240}.
\newblock
\urldef\tempurl%
\url{https://doi.org/10.1007/978-3-319-41528-4\_12}
\showDOI{\tempurl}


\bibitem[\protect\citeauthoryear{Tuerk}{Tuerk}{2012}]%
        {Tuerk-NFA}
\bibfield{author}{\bibinfo{person}{Thomas Tuerk}.}
  \bibinfo{year}{2012}\natexlab{}.
\newblock \bibinfo{title}{{A Formalisation of Finite Automata in Isabelle /
  HOL}}.
\newblock
  \bibinfo{howpublished}{\url{https://www.thomas-tuerk.de/assets/talks/cava.pdf}}.
\newblock


\bibitem[\protect\citeauthoryear{Veanes, Hooimeijer, Livshits, Molnar, and
  Bj{\o}rner}{Veanes et~al\mbox{.}}{2012}]%
        {symbolic-transducer}
\bibfield{author}{\bibinfo{person}{Margus Veanes}, \bibinfo{person}{Pieter
  Hooimeijer}, \bibinfo{person}{Benjamin Livshits}, \bibinfo{person}{David
  Molnar}, {and} \bibinfo{person}{Nikolaj Bj{\o}rner}.}
  \bibinfo{year}{2012}\natexlab{}.
\newblock \showarticletitle{Symbolic finite state transducers: algorithms and
  applications}. In \bibinfo{booktitle}{\emph{Proceedings of the 39th {ACM}
  {SIGPLAN-SIGACT} Symposium on Principles of Programming Languages, {POPL}
  2012, Philadelphia, Pennsylvania, USA, January 22-28, 2012}},
  \bibfield{editor}{\bibinfo{person}{John Field} {and} \bibinfo{person}{Michael
  Hicks}} (Eds.). \bibinfo{publisher}{{ACM}}, \bibinfo{pages}{137--150}.
\newblock
\urldef\tempurl%
\url{https://doi.org/10.1145/2103656.2103674}
\showDOI{\tempurl}


\bibitem[\protect\citeauthoryear{Wang, Tsai, Lin, Yu, and Jiang}{Wang
  et~al\mbox{.}}{2016}]%
        {SLOG}
\bibfield{author}{\bibinfo{person}{Hung{-}En Wang},
  \bibinfo{person}{Tzung{-}Lin Tsai}, \bibinfo{person}{Chun{-}Han Lin},
  \bibinfo{person}{Fang Yu}, {and} \bibinfo{person}{Jie{-}Hong~R. Jiang}.}
  \bibinfo{year}{2016}\natexlab{}.
\newblock \showarticletitle{String Analysis via Automata Manipulation with
  Logic Circuit Representation}. In \bibinfo{booktitle}{\emph{Computer Aided
  Verification - 28th International Conference, {CAV} 2016, Toronto, ON,
  Canada, July 17-23, 2016, Proceedings, Part {I}}}
  \emph{(\bibinfo{series}{Lecture Notes in Computer Science},
  Vol.~\bibinfo{volume}{9779})}, \bibfield{editor}{\bibinfo{person}{Swarat
  Chaudhuri} {and} \bibinfo{person}{Azadeh Farzan}} (Eds.).
  \bibinfo{publisher}{Springer}, \bibinfo{pages}{241--260}.
\newblock
\urldef\tempurl%
\url{https://doi.org/10.1007/978-3-319-41528-4\_13}
\showDOI{\tempurl}


\bibitem[\protect\citeauthoryear{Yu, Alkhalaf, Bultan, and Ibarra}{Yu
  et~al\mbox{.}}{2014}]%
        {DBLP:journals/fmsd/YuABI14}
\bibfield{author}{\bibinfo{person}{Fang Yu}, \bibinfo{person}{Muath Alkhalaf},
  \bibinfo{person}{Tevfik Bultan}, {and} \bibinfo{person}{Oscar~H. Ibarra}.}
  \bibinfo{year}{2014}\natexlab{}.
\newblock \showarticletitle{Automata-based symbolic string analysis for
  vulnerability detection}.
\newblock \bibinfo{journal}{\emph{Formal Methods Syst. Des.}}
  \bibinfo{volume}{44}, \bibinfo{number}{1} (\bibinfo{year}{2014}),
  \bibinfo{pages}{44--70}.
\newblock
\urldef\tempurl%
\url{https://doi.org/10.1007/s10703-013-0189-1}
\showDOI{\tempurl}


\end{thebibliography}
%

\end{document}